\documentclass[numberedappendix]{emulateapj}

\usepackage{graphicx}% Include figure files
\usepackage{dcolumn}% Align table columns on decimal point
\usepackage{bm}% bold math
\usepackage{epsf}
\usepackage{verbatim}
\usepackage{amsmath}
\usepackage{amsfonts}
\usepackage{ifthen}
\newboolean{incApp}
\setboolean{incApp}{false}
\newcommand{\mycond}[2]{\ifthenelse{#1}{#2}{}}
\newcommand{\myNcond}[2]{\ifthenelse{#1}{}{#2}}
\newcommand{\till}{{\mbox{--}}}
\newcommand{\myMomentum}{{\Pi}}
\newcommand{\myv}{{v}}

\newcommand{\mywA}{{w_1}}
\newcommand{\mywAA}{{w_1^2}}
\newcommand{\mywB}{{w_2}}
\newcommand{\myWA}{{w_1}}
\newcommand{\myWAA}{{w_1^2}}
\newcommand{\myWB}{{w_2}}
\newcommand{\myx}{{\zeta}}
\newcommand{\mysigma}{{\sigma}}
\newcommand{\myenergy}{{E}}
\newcommand{\myF}{{F}}
\newcommand{\myQ}{{\mathcal{Q}}}
\newcommand{\myP}{{\mathcal{P}}}
\newcommand{\myU}{{\mathcal{U}}}
\newcommand{\myfa}{{g_1}}
\newcommand{\myfb}{{g_2}}
\newcommand{\myE}{{\mathcal{E}}}
\newcommand{\myQb}{{\mathbb{Q}}}
\newcommand{\myPb}{{\mathbb{P}}}
\newcommand{\myUb}{{\mathbb{U}}}
\newcommand{\myV}{{\mathbb{V}}}

\newcommand{\fig}[1]{{Figure \ref{#1}}}
\def\myfig#1{./#1}

\newcommand{\BoldA}[1]{{#1}}

\newcommand{\ie}{\emph{i.e.} }
\newcommand{\eg}{\emph{e.g.,} }

\newcommand{\be}{\begin{equation}}
\newcommand{\ee}{\end{equation}}
\newcommand{\bea}{\begin{equation*}}
\newcommand{\eea}{\end{equation*}}
\newcommand{\beqr}{\begin{eqnarray} \nonumber}
\newcommand{\eeqr}{\end{eqnarray}}
\newcommand{\beqrb}{\begin{eqnarray}}
\newcommand{\eeqrb}{\nonumber \end{eqnarray}}
\newcommand{\fin}{\mbox{ .}}
\newcommand{\coma}{\mbox{ ,}}

\newcommand{\const}{\mbox{const.}}

\newcommand{\gama}{$\gamma$}

\newcommand{\unit}[1]{\bm{\hat{#1}}}
\newcommand{\pr}{\partial}

\newcommand{\myNi}{\emph{(i)}\,}
\newcommand{\myNii}{\emph{(ii)}\,}
\newcommand{\myNiii}{\emph{(iii)}\,}
\newcommand{\myNiv}{\emph{(iv)}\,}
\newcommand{\myNv}{\emph{(v)}\,}

\shorttitle{Self-similar jet}
\shortauthors{Keshet \& Kogan}

\begin{document}

\title{Self-similar ultra-relativistic jetted blast wave}

\author{
Uri Keshet
and
Dani Kogan
}

\email{ukeshet@bgu.ac.il}

\affiliation{
Physics Department, Ben-Gurion University of the Negev, P.O. Box 653, Be'er-Sheva 84105, Israel
}

\begin{abstract}
Following a suggestion that a directed relativistic explosion may have a universal intermediate asymptotic, we derive a self-similar solution for an ultra-relativistic jetted blast wave.
The solution involves three distinct regions: an approximately paraboloid head where the Lorentz factor $\gamma$ exceeds $\sim1/2$ of its maximal, nose value; a geometrically self-similar, expanding envelope slightly narrower than a paraboloid; and an axial core in which the \BoldA{(cylindrically, henceforth)} radial flow \BoldA{$u$} converges inward toward the axis.
Most ($\sim 80\%$) of the energy lies well beyond the \BoldA{leading,} head \BoldA{region}.
Here, a radial cross section shows a maximal $\gamma$ (separating the core and the envelope), a sign reversal in \BoldA{$u$}, and a minimal $\gamma$, at respectively $\sim 1/6$, $\sim1/4$, and $\sim3/4$ of the shock radius.
The solution is apparently unique, and approximately agrees with previous simulations, of different initial conditions, that resolved the head.
This suggests that unlike a spherical relativistic blast wave, our solution is an attractor, and may thus describe directed blast waves such as in the external shock phase of a $\gamma$-ray burst.
\end{abstract}


\maketitle

\section{Introduction}
\label{sec:Intro}

Relativistic jets give rise to some of the most spectacular astronomical sources, including \gama-ray bursts (GRB) with Lorentz factors $\Gamma\sim 10^{2}\till10^{3}$, active galactic nuclei (AGN) with $\Gamma\sim 10$, and microquasars with $\Gamma\sim$ a few.
Such jets are also likely to form in systems currently invisible to us, such as failed supernova explosions.
Yet, there is \BoldA{considerable} uncertainty regarding the origin, structure, and evolution of relativistic jets.

One might expect that in the limit of an evolved, ultra-relativistic jetted blast wave (henceforth jet), propagating into a sufficiently weakly magnetized, homogeneous medium, the lack of a characteristic length scale would lead the structure of the jet to approach some self-similar attractor. Such a solution may provide a framework for studying jets, and be useful as an approximate description of directed blast waves in astronomical systems, such as in the external shocks stage of a GRB.

It has been argued \citep[][henceforth G07]{Gruzinov07} that a directed explosion, in which the total momentum $\myMomentum$ is comparable \BoldA{(when multiplied by the speed of light $c$)} to the total energy $\myE_{tot}$, such that $\myE_{tot}-\myMomentum\, c\ll \myE_{tot}$, may have a well-defined self-similar attractor.
This differs from a non-directed ($\myE_{tot}-\myMomentum \,c\lesssim \myE_{tot}$) explosion, which is not in full causal contact when highly relativistic, and thus has no universal intermediate asymptotic.
Indeed, non-directed explosions asymptote to the \citet{BlandfordMcKee76} self-similar solution in the ultra-relativistic phase only if they are initially spherically symmetric \citep{Gruzinov00}.

The structure of a relativistic jet was described qualitatively in \citet{Rhoads99}, and quantitatively in G07, where numerical simulations of various initial conditions were argued to approach a unique attractor solution.
The \BoldA{reported} jet has a nearly \BoldA{paraboloidal} shock front, and, with increasing distance from the shock, shows a monotonic decline in the local Lorentz factor $\gamma$, the proper pressure $p$, and the \BoldA{(cylindrically, henceforth)} radial velocity $u$.

While these results are promising, the universality, uniqueness, and full structure of the attractor were until now unclear.
The jet structure was derived from simulations rather than by \BoldA{directly} solving the flow equations.
This structure was reported only for the head region, within a distance $1<\xi\lesssim5$ of the light signal preceding the jet, normalized such that $\xi=1$ is the nose position \BoldA{(see Eq.~(\ref{eq:SeS_coord}) for a precise definition of $\xi$)}.
The simulations were reported to approach the attractor very slowly, suggesting that they have not fully converged \BoldA{upon it}.
A possible sign reversal in $u$ was reported far downstream, probably in the non-converged region, suggesting some deviation from a parabolic profile (G07).

Several questions have thus remained open. What are the full equations describing the self-similar, ultra-relativistic jet? These equations were only partly derived in G07. What are the solution or solutions to these equations? What is the corresponding normalization of the self-similarity scaling? Is the solution unique, and if so - what provides closure to the equation system? What is the structure of the jet, in particular far from the nose? Does it show the flow monotonicity reported at the head? Is most of the energy contained in the head, or does it lie beyond it, in regions not yet reported, and probably not converged?

Here we address these questions.
First, we derive the equations for an ultra-relativistic, self-similar jet in \S\ref{sec:equations}, following and supplementing G07.
Analytic and semi-analytic constraints on the solution are derived in \S\ref{sec:Analytic}.
In \S\ref{sec:NumericalSolution}, we \BoldA{finally} solve the equations numerically, and show that their regular solution is unique \BoldA{and satisfies the constraints obtained in \S\ref{sec:Analytic}}.
The results are summarized and discussed in \S\ref{sec:Discussion}.
The observational implications and stability of the solution are deferred to a forthcoming publication.

We assume \myNi an axisymmetric relativistic flow expanding into a homogeneous medium in flat space; \myNii that the effects of cooling, electromagnetic fields, and self-gravity are negligible; and \myNiii self-similarity in time.
We normalize the speed of light, $c=1$.

\section{Self-similar ultra-relativistic equations}
\label{sec:equations}

\subsection{Hydrodynamic equations}

Following G07 and using similar notations, we consider a directed flow propagating along the positive $z$ direction, where $x^\mu=(t,x,y,z)$ are cartesian coordinates in the upstream frame.
The flow \BoldA{downstream of the shock} is parameterized by the squared Lorentz factor $q\equiv \gamma^2$, the proper pressure $p$, and the cylindrically radial (\ie perpendicular to the axis of symmetry) velocity $u$.
We thus analyze these downstream properties using upstream frame coordinates, henceforth omitting the terms upstream and downstream in this context.

The flow is governed by the relativistic hydrodynamic equation
\begin{equation} \label{Eq:Tmunu}
\pr_\nu T^{\mu\nu} = 0 \coma
\end{equation}
where $T^{\mu\nu} = (4u^\mu u^\nu-g^{\mu\nu})p$ is the energy-momentum tensor, $u^\mu$ is the four-velocity, $g^{\mu\nu}$ is the Minkowski metric\BoldA{, and we assumed a relativistic fluid with an enthalpy density of $4p$}.
Assuming axial symmetry in cylindrical coordinates $(t,r,\phi,z)$, the four-velocity becomes $u^\mu=\gamma(1,u\unit{r}+\myv\unit{z})$. The $t$, $z$ and $r$ components of Eq.~(\ref{Eq:Tmunu}) are then
\begin{equation} \label{eq:hydrot}
\pr_t\left[(4q-1)p\right]+4\pr_z(qp\myv )+4r^{-1}\pr_r(rqpu)=0 \coma
\end{equation}
\begin{equation} \label{eq:hydroz}
4\pr_t(qp\myv )+\pr_z\left[(4q\myv ^2+1)p\right]+4r^{-1}\pr_r(rqpu\myv )=0 \coma
\end{equation}
and
\begin{equation} \label{eq:hydror}
4\pr_t(qpu)+4\pr_z(qpu\myv )+4r^{-1}\pr_r(rqpu^2)+\pr_rp=0 \fin
\end{equation}

In the upstream medium (denoted by a tilde), pressure is negligible, so the only non-zero component in its energy-momentum tensor $\tilde{T}^{\mu\nu}$ is $\tilde{e}\equiv \tilde{T}^{00}$.
Parameterizing the shock \BoldA{(subscript $s$, henceforth)} surface as the zero isosurface of a scalar function
\begin{equation} \label{eq:zs_of_r}
S(x^\mu)\equiv z-z_s(t,r)=0 \coma
\end{equation}
the requirement of continuous energy-momentum fluxes across the shock can be written as
\begin{equation} \label{eq:taumunu}
T^{\mu\nu}\pr_\nu S = \tilde{T}^{\mu\nu}\pr_\nu S \fin
\end{equation}
The $t$, $z$ and $r$ components of these constraints become
\begin{equation}\label{eq:BCt}
4q_\BoldA{s}p_\BoldA{s}(\pr_tz_s-\myv_\BoldA{s} +u_\BoldA{s}\pr_rz_s)-p\pr_tz_s = \tilde{e}\pr_tz_s \coma
\end{equation}
\begin{equation} \label{eq:BCz}
4q_\BoldA{s}\myv_\BoldA{s} (\pr_tz_s-\myv_\BoldA{s} +u_\BoldA{s}\pr_rz_s) - 1 = 0 \coma
\end{equation}
and
\begin{equation} \label{eq:BCr}
4q_\BoldA{s}u_\BoldA{s}(\pr_tz_s-\myv_\BoldA{s} +u_\BoldA{s}\pr_rz_s)+\pr_rz_s = 0 \fin
\end{equation}

The flow is given by a solution to equations (\ref{eq:hydrot}--\ref{eq:hydror}), with shock boundary conditions (\ref{eq:BCt}--\ref{eq:BCr}).
Notice that in Eq.~(\ref{eq:zs_of_r}) we assumed that $z_s$ is a function of $r_\BoldA{s}$, and not vice versa, such that the jet is infinitely long, and monotonically widens with $z$.
This excludes possible scenarios in which the jet widens near the head but narrows down farther downstream, as found in some simulations \citep[\eg][]{Granot_et_al_00, Cannizzo_et_al_04}. The assumption can be avoided by parameterizing the shock using $r_s(t,z)$, instead of $z_s$; see \S\ref{sec:Discussion}.

\subsection{Ultra-relativistic limit}

In the ultra-relativistic limit, $q^{-1}$ can be used as a small expansion parameter, giving
\begin{equation}
\myv = \sqrt{1-q^{-1}-u^2} = 1-\frac{1+qu^2}{2q} +O\left(q^{-2}\right) \coma
\end{equation}
where we assumed the scaling $qu^2=O(1)$, as confirmed by the results (see G07 and \S\ref{sec:NumericalSolution}).
We define
\BoldA{
%$\mywA=1+qu^2$ and $\mywB=1+2qu^2$,
\begin{equation} \label{eq:Defs_w1_w2}
\mywA\equiv 1+qu^2 \quad \mbox{and} \quad \mywB\equiv 1+2qu^2 \coma
\end{equation}
}
for future convenience.

It is useful to replace $z$ by a coordinate $\myx\equiv t-z$, measuring the distance to a plane \BoldA{perpendicular to the jet and} moving ahead of \BoldA{it} at the speed of light.
Equations (\ref{eq:hydrot}\till\ref{eq:hydror}) then become, to leading order in $q^{-1}$,
\begin{equation}\label{eq:URt}
4\pr_t(qp)+\pr_\myx(\mywB p) + (4/r)\pr_r(rqpu) = 0 \coma
\end{equation}
\begin{equation}\label{eq:URz}
\pr_t(\mywB p)+\pr_\myx(\mywAA p/q) + (2/r)\pr_r(r\mywA pu) = 0 \coma
\end{equation}
and
\begin{equation}\label{eq:URr}
4qp\pr_tu+2p\mywA\pr_\myx u+u\pr_\myx p +\pr_rp + 4qpu\pr_ru = 0 \fin
\end{equation}
Note that we added the last term in Eq.~(\ref{eq:URr}) to correct equation (18) of G07.

In a similar fashion, the shock boundary conditions \BoldA{(\ref{eq:BCt}--\ref{eq:BCr})} at $\myx=\myx_s(t,r)\equiv t-z_s$ become, to leading order in $q^{-1}$,
\begin{equation} \label{eq:URBC}
q_\BoldA{s}=\frac{1}{4\pr_t\myx_s+2(\pr_r\myx_s)^2} \coma \quad
p_\BoldA{s}=\frac{4}{3}\tilde{e}q_\BoldA{s} \coma \quad
\mbox{and} \quad
u_\BoldA{s}=\pr_r \myx_s \fin
\end{equation}

\subsection{Self-similarity}

The ultra-relativistic Eqs.~(\ref{eq:URt}\till\ref{eq:URBC})
remain invariant if the vector
\begin{equation} \label{eq:PowerLawScaling}
\myV \equiv \left( \myx t, r^2, \myx^2 q, \myx^2 p, \myx^2/u^2 \right)
\end{equation}
is multiplied by an arbitrary constant, \BoldA{and so they} admit a self-similar scaling.
In $D$ spatial dimensions, the total energy of a self-similar solution would scale as
\begin{equation}\label{eq:EnergyScaling}
\myE_{tot} \propto \myx_* r_*^{D-1}p_*q_*\propto \myx_*^{\frac{D-3}{2}}t^{\frac{D+3}{2}} \coma
\end{equation}
where a star denotes a typical value at time $t$.
This energy is constant during the self-similar phase, imposing scalings such as $\myx_*\propto t^{\BoldA{-}(D+3)/(D-3)}$ and $r_*\propto t^{\BoldA{-}3/(D-3)}$.
These scalings hold for $D\neq3$, but in three spatial dimensions they degenerate into exponential functions.
Indeed, the equation system remains invariant under rescaling of the vector $\myV$ in Eq.~(\ref{eq:PowerLawScaling}), even if \BoldA{the first,} $t$\BoldA{-dependent component} is eliminated from $\myV$.

The \BoldA{resulting} exponential scaling motivates the dimensionless, self-similar, \BoldA{respectively axial and radial} coordinates
\begin{equation} \label{eq:SeS_coord}
\xi \equiv \Lambda^2 \frac{\zeta}{\tau}
\quad \mbox{and} \quad
\eta \equiv \Lambda  \frac{r}{\tau} \coma
\end{equation}
and the self-similar functions \BoldA{describing respectively the Lorentz factor squared, the pressure, and the radial velocity,}
\begin{equation} \label{eq:SeS_vars}
Q \equiv \Lambda^{-2} q \coma \quad
P \equiv \Lambda^{-2} \frac{p}{\tilde{e}} \coma \quad
\mbox{and} \quad
U \equiv \Lambda u \fin
\end{equation}
Here, the \BoldA{temporal} scaling factor
\begin{equation}
\Lambda \equiv e^{-\frac{t-t_f}{\tau}}
\end{equation}
is large in the self-similar regime,
\begin{equation} \label{eq:SS_tau}
\tau =
C\left(\frac{\myE_{tot}}{\tilde{e}} \right)^{1/3}
\end{equation}
is the e-fold expansion time, \BoldA{$C$ is a dimensionless constant,} and we defined a final time $t_f$ when the flow is no longer relativistic.
The approximation holds for $q\gg 1$
and
$u\ll 1$, so $(t-t_f)$ is assumed negative and large \BoldA{with respect to $\tau$}.

The total energy in the jet is related by $\myE_{tot}=\tau^3 \tilde{e}\myenergy_{tot}$
to the total self-similar energy
\begin{equation} \label{eq:SS_energy}
\myenergy_{tot}
= \int \myenergy \, dV
= 8\pi \int_0^\infty \eta \,d\eta \int_{\xi_s(\eta)}^\infty d\xi \, QP
\coma
\end{equation}
where the self-similar energy density is $\myenergy=4QP$.
This fixes the constant $C=\myenergy_{tot}^{-1/3}$.

Rewriting Eqs.~(\ref{eq:URt}\till\ref{eq:URBC}) using these definitions finally gives
the $t$, $z$ and $r$ components of Eq.~(\ref{Eq:Tmunu}) in the self-similar regime,
\begin{equation}\label{eq:SSt}
(4+\pr_\psi)(QP)-\frac{\pr_\xi(\myWB P)}{4}-\frac{\pr_\eta(\eta QPU)}{\eta}=0 \coma
\end{equation}
\begin{equation}\label{eq:SSz}
(2+\pr_\psi)(\myWB P)-\pr_\xi(\myWAA P/Q)-\frac{2\pr_\eta(\eta\myWA PU)}{\eta}=0 \coma
\end{equation}
and
\begin{equation}\label{eq:SSr}
(U\pr_\eta+1-\pr_\psi)U+\frac{\myWA}{2Q}\pr_\xi U+\frac{\pr_\eta+U\pr_\xi}{4QP}P=0 \coma
\end{equation}
where we defined $\psi=\log(\xi^{1/2}\eta)$ for brevity.
The boundary conditions are written in self-similar form on \BoldA{the surface} $\xi=\xi_s(\eta)\equiv (\Lambda^2/\tau)\myx_s$, as
\begin{equation}\label{eq:SSBC}
Q_\BoldA{s}=\frac{1}{8\xi_s-4\eta\xi_s'+2\xi_s'^2} \coma \quad P_\BoldA{s}=\frac{4}{3}Q_\BoldA{s} \coma \quad \mbox{ and}\quad U_\BoldA{s}=\xi_s' \fin
\end{equation}
The scaling (\ref{eq:SeS_vars}) implies that $\myWA=1+QU^2$ and $\myWB=1+2QU^2$ can be written in a self-similar fashion\BoldA{, in a form identical to Eq.~(\ref{eq:Defs_w1_w2})}.
Note that we added the first term in \BoldA{the parenthesis of} Eq.~(\ref{eq:SSr}) to correct equation 27 in G07.

\subsection{Equation properties}

The self-similar Eqs.~(\ref{eq:SSt}--\ref{eq:SSr}) and boundary conditions (\ref{eq:SSBC}) \BoldA{remain} invariant if $\Lambda$ is multiplied by an arbitrary constant.
This constant can in principle be chosen as $(-1)$, indicating that $Q$ and $P$ are symmetric, while $U$ is antisymmetric, under \BoldA{the} reflection \BoldA{$\eta\to (-\eta)$} across the \BoldA{axis of} symmetry (henceforth, the axis).

Without loss of generality, henceforth we choose \BoldA{this arbitrary} constant such that the \BoldA{nose, \ie the very} head \BoldA{(subscript $h$, henceforth)} of the jet is at
\begin{equation}
\xi = \xi_h\equiv\xi_s(\eta=0)=1 \fin
\end{equation}
Equation (\ref{eq:SSBC}) then implies that at the head, the flow parameters are given by
\begin{equation}
Q_h = \frac{1}{8} \coma \quad
P_h = \frac{1}{6} \coma \quad
U_h = 0 \coma \quad
\mbox{and} \quad
E_h = \frac{1}{12} \fin
\end{equation}
Here and below, we define $\myF_s$, $\myF_a$ and $\myF_h$ as the field $\myF\in\{Q,P,U,E\}$ evaluated at the shock, on the axis, and at the head, respectively.

Notice that Eqs.~(\ref{eq:SSt}--\ref{eq:SSr}) are, in addition, unchanged if $P$ is multiplied by an arbitrary constant, but this constant is fixed by the boundary conditions (\ref{eq:SSBC}).

The equation system involves three first-order partial differential equations, along with three boundary conditions, for the three \BoldA{fields} $\{Q,P,U\}$ living in the two dimensional $\xi$--$\eta$ space. In addition, the equation system depends on the unknown one-dimensional function $\xi_s(\eta)$, so one additional constraint may appear to be missing.
As we show below, the system is closed, and the shock profile $\xi_s(\eta)$ is fixed, by requiring that the solution be regular, even without specifying boundary conditions far downstream.

The oriented derivative along the shock may be defined as
\begin{equation} \label{eq:ShockDerivative}
\partial_s \equiv \partial_\eta+\xi_s'(\eta)\partial_\xi \fin
\end{equation}
\BoldA{For a shock profile that is monotonic, in the sense that the function $\xi_s(\eta)$ monotonically increases, }
%The shock profile $\xi_s(\eta)$ of an infinite jet is monotonic, so
one may alternatively write the derivative \BoldA{in Eq.~(\ref{eq:ShockDerivative})} and the boundary conditions (\ref{eq:SSBC}) using \BoldA{the parametrization} $\eta_s(\xi)$, instead of $\xi_s(\eta)$.
This more general parametrization is used in \S\ref{sec:NumericalSolution}.

If the shock profile is known, or is postulated, one may use Eq.~(\ref{eq:ShockDerivative}) to turn the self-similar Eqs.~(\ref{eq:SSt}--\ref{eq:SSBC}) into an infinite series of ordinary differential equations for increasingly high-order field derivatives, in either the $\xi$ or the $\eta$ direction.
This is utilized below, in \S\ref{sec:Analytic} and in \S\ref{sec:HeadModel}.

For a \BoldA{monotonic} shock profile, $\xi_s'(\eta)>0$\BoldA{, Eq.}
(\ref{eq:SSBC}) implies that $U_s>0$, \ie near the shock, the radial flow is \BoldA{always} directed outwards.
Combining Eqs.~(\ref{eq:SSr}), (\ref{eq:SSBC}) and (\ref{eq:ShockDerivative}) indicates that
%the radial acceleration vanishes just behind the shock,
\begin{equation} \label{eq:NoRAcceleration}
\partial_\xi U_s = 0 \coma
\end{equation}
\BoldA{which may be interpreted as a vanishing self-similar radial acceleration.}
Note that the term we added to equation 27 of G07, in order to obtain Eq.~(\ref{eq:SSr}), is essential for recovering this property.
In particular it implies, using Eq.~(\ref{eq:ShockDerivative}), that $\partial_\eta U_s = \xi_s''$.

\section{Semi-analytic constraints}
\label{sec:Analytic}

\subsection{Overview of the solution}
\label{sec:AnalyticGeneral}

Expand the shock profile about the nose of the jet, $\xi_s=1+\xi_2\eta^2+\xi_4\eta^4+O(\eta^6)$, where \BoldA{$\xi_2$ and $\xi_4$ are numerical constants, and} we used reflection symmetry to omit odd powers of $\eta$.
The coefficient $\xi_2$ is unlikely to vanish, as confirmed numerically below, so the head of the jet is approximately a paraboloid.
Define the logarithmic shock profile slope,
\begin{equation}
\beta \equiv \frac{d \ln \xi_s}{d\ln \eta} \coma
\end{equation}
such that at the nose $\beta\to 2$.

A parabolic, $\beta=2$ shock profile is stationary, in the sense that a point $\{r,\myx_s(r)\propto r^2\}$
is mapped at a time $\Delta t$ later onto $\{\Lambda r, \Lambda^2 \myx_s \propto (\Lambda r)^2 \}$, with $\Lambda = e^{\Delta t/\tau}$.
In non self-similar coordinates, a self-similar $\beta<2$ (\ie wide jet) profile narrows down in time, eventually approaching $\myx_s\sim r^2$, whereas a $\beta>2$ (narrow) shock gradually widens toward $\myx_s\sim r^2$.

Far from the head, the self-similar shock profile must become narrower than a paraboloid, \ie $\beta>2$, because $\beta\leq2$ leads to nonphysical results, in particular a divergence of the energy in Eq.~(\ref{eq:SS_energy}). Moreover, $\beta<2$ leads to only non-real valued solutions (see \S\ref{sec:NoBetaLess2}), while $\beta=2$ leads to a divergence of $Q$ (see \S\ref{sec:beta2} and \fig{fig:GSS_beta2}), and cannot be matched to the axial region (see \S\ref{sec:NoBetaEq2}).

At large distances from the head, one may approximate $\beta$ as a constant, up to logarithmic corrections.
For such power-law behavior, $\xi_s\propto \eta^\beta$, an additional, geometrical self-similarity (GSS) in the $\xi$--$\eta$ plane may emerge far from the head and from the axis, with the self similar parameter (G07)
\begin{equation} \label{eq:SS_parameter}
\chi_\BoldA{0}\equiv \xi/\eta^\beta \coma
\end{equation}
as discussed in \S\ref{sec:GSSequations}.

A GSS scaling in the relevant, $\beta>2$ regime (demonstrated in \S\ref{sec:beta202} and in \fig{fig:Beta202}) breaks down near the head and near the axis, as \BoldA{equation} terms that do not follow the GSS scaling become large and can no longer be neglected.
Denote the surface along which the axial terms break the GSS scaling as $\xi=\xi_c(\eta)$.
One may define this surface as the boundary between an axial, or core, region at $\xi>\xi_c$, and the GSS region at $\xi<\xi_c$.
In \S\ref{sec:NumericalSolution} we show numerically that this boundary can be associated with a maximal value of $Q(\eta)$, giving $\xi_c\simeq 20\eta^\beta$.

In our $\beta>2$ regime, Eqs.~(\ref{eq:SSBC}--\ref{eq:NoRAcceleration}) imply that $\partial_\eta U_s=\xi_s''>0$, so near the shock the (positive) radial velocity increases radially outwards.
At large radii, $\beta>2$ implies that $Q$ and $P$ similarly become monotonically larger as one approaches the shock, such that $\pr_\xi \{Q_s,P_s\}<0$ and $\pr_\eta\{Q_s,P_s\}>0$, as seen by solving Eqs.~(\ref{eq:SSt}--\ref{eq:SSBC}), with the aid of Eq.~(\ref{eq:ShockDerivative}), for these derivatives at the shock.

This behavior, in which $\myF\in\{Q,P,U,E\}>0$ increases monotonically toward the shock, namely
\begin{equation}
\pr_\xi \myF\leq0 \quad \mbox{and} \quad \pr_\eta \myF>0\coma
\end{equation}
is henceforth referred to as flow monotonicity.
The simulations of G07 show this type of behavior throughout the reported, $1\leq\xi<5$ regime.
It is therefore natural to ask if flow monotonicity persists throughout the jet.

Within the GSS regime, $\beta>2$ solutions are found to transition at some $\chi_\BoldA{0}=\chi_u$, from $U(\chi_\BoldA{0}<\chi_u)>0$ toward the shock, to $U(\chi_\BoldA{0}>\chi_u)<0$ toward the axis (see \S\ref{sec:beta202}).
Indeed, far from the head, $Q_a(\xi\gg\xi_h)\propto\xi^{-\alpha}$ approaches a power-law along the axis, with $\alpha\leq1$, \BoldA{which, as we show, implies} that $\pr_\eta U(\eta=0)\to(\alpha-1)$ asymptotes to a negative constant (except in the special case $\alpha=1$; see \S\ref{sec:Axis}).
This is consistent with a $U<0$ inflow emanating at $\chi_u$ within the GSS regime, and extending all the way to the axis.
Thus, the radial velocity reverses sign, corresponding to a radial inflow near the axis, and a radial outflow near the shock.
As the \BoldA{radial} velocity vanishes along the axis, $U_a=0$, this indicates that $U$ cannot be monotonic.

To see that $Q$ cannot be monotonic either, recall that along the axis, $Q_a\propto\xi^{-\alpha}$ declines with increasing $\xi$ no faster than $\xi^{-1}$.
In contrast, along the shock, Eq.~(\ref{eq:SSBC}) with $\beta>2$ implies that $Q_s\propto\xi^{-2(1-\beta^{-1})}$ declines faster than $\xi^{-1}$.
Hence, there must be a region where $Q$ declines as $\eta$ increases toward the shock, ruling out monotonicity in $Q$.
This behavior is confirmed in the GSS regime (see \S\ref{sec:beta2} and \S\ref{sec:beta202}).

Nevertheless, flow monotonicity does manifest in the head region. Requiring such monotonicity near the head implies that the jet cannot be too narrow, constrains the flow in the head region, and limits the extent of this region to $\xi<\xi_g\simeq 5$ (see \S\ref{sec:MonotonicHead} and \fig{fig:HeadMono}).

Although the shock profile is not directly constrained by the system of equations, and no far downstream boundary condition is imposed, \BoldA{we find that} only a unique profile avoids a divergence of the flow.
For example, if the jet is too wide, $q=\gamma^2$ becomes negative on the axis (see \S\ref{sec:WideJets}), which is both nonphysical and leads to divergencies.
Indeed, as shown in \S\ref{sec:NumericalSolution}, numerically solving the equations and requiring a regular jet picks out a unique flow solution (see Figures \ref{fig:ParameterScan}--\ref{fig:NumericJet}), which agrees with all the features mentioned above and derived quantitatively below.

The above arguments indicate that a regular self-similar jet has a unique solution, composed of three distinct flow regions: a monotonic head region ($\xi\lesssim\xi_g$), an effectively one-dimensional, GSS envelope ($\xi_g\lesssim\xi\lesssim\xi_c$), and an axial, or core, region ($\xi\gtrsim\xi_c$). A radial inflow encompasses the core and the \BoldA{inner} envelope regimes.
Only the head and the outer part of the envelope\BoldA{, which harbor a radial outflow,} show a monotonic flow.

\subsection{Geometric self-similarity}
\label{sec:GSSequations}

At large distances from the head of the jet, the shock profile may be approximated by a power law, in which case the temporally self-similar Eqs.~(\ref{eq:SSt}--\ref{eq:SSBC}) may be further simplified.
Using \BoldA{a} scaling parameter $\chi_\BoldA{0}$ \BoldA{such as defined in} Eq.~(\ref{eq:SS_parameter}), these equations can be cast in a geometrically\BoldA{, and not only temporally,} self-similar form.

It is useful to introduce a slightly different GSS parameter,
\begin{equation} \label{eq:SS_parameterB}
\chi=\frac{\xi-\xi_h}{A\eta^\beta} \coma
\end{equation}
where the unknown normalization constant $A$ is \BoldA{introduced} such that the position of the shock in the GSS regime can be taken as $\chi=\chi_s$, with a constant $\chi_s$ which we choose as $\chi_s\equiv1$.

The case $\beta<2$ does not lead to physical GSS solutions, as \BoldA{mentioned above and} discussed in \S\ref{sec:NoBetaLess2}.
Therefore, here we focus on $\beta\geq2$.

The flow equations can then be written approximately as functions of $\chi$, if the parameters are rescaled as
\begin{widetext}
\begin{equation}\label{eq:GSS_scaling}
\myQ(\chi) \equiv A^2\eta^{2(\beta-1)}Q \coma \quad
\myP(\chi) \equiv A^2\eta^{2(\beta-1)}P \coma \quad \mbox{and} \quad
\myU(\chi) \equiv A^{-1}\eta^{-(\beta-1)}U \fin
\end{equation}
Equations (\ref{eq:SSt}--\ref{eq:SSr}) now become, respectively,
\begin{equation} \label{eq:GSS_t}
\frac{(\myWB\myP)'}{4}- (3\beta-4)\myQ \myP \myU -\beta \chi (\myQ \myP \myU )'
- \frac{2\xi_h}{\eta^{2(\beta-1)}A^2}(\myQ \myP )' +\frac{\beta-2}{\eta^{\beta-2}A}[(\chi \myQ \myP )'+3\myQ \myP ]=0
\coma
\end{equation}
\begin{equation} \label{eq:GSS_z}
\left(\frac{\myWAA\myP}{\myQ}\right)'-2\beta \chi (\myWA\myP \myU)'
-2(\beta-2)\myWA\myP \myU  -\frac{2\xi_h}{\eta^{2(\beta-1)}A^2} (\myWB\myP)'
+ \frac{\beta-2}{\eta^{\beta-2}A\chi} (\chi^2\myWB\myP)' =0
\coma
\end{equation}
and
\begin{equation} \label{eq:GSS_r}
(\myWA -2\beta \chi  \myQ \myU)\myU' - (2-\myWB)(\beta-1) + (\myU -\beta \chi )\frac{\myP'}{2\myP}
-\frac{4\xi_h}{\eta^{2(\beta-1)}A^2}\myQ \myU'
+\frac{2(\beta-2)\chi^2 \myQ}{\eta^{\beta-2}A} \left(\frac{\myU}{\chi}\right)' =0\coma
\end{equation}
\end{widetext}
where the scaling (\ref{eq:GSS_scaling}) maintains $\myWA=1+\myQ\myU^2$ and $\myWB=1+2\myQ\myU^2$ in GSS form.
Note that in Eq.~(\ref{eq:GSS_r}), we added the second parenthesis, and the second term in the first parenthesis, to correct equation 36 of G07.

The last two terms in each of the above three equations (\ref{eq:GSS_t}--\ref{eq:GSS_r}) do not follow the GSS scaling.
However, the first of the two terms is negligible far from the head, and the second is negligible far from the axis.
Note that the last term in each equation vanishes in the special case $\beta=2$.
GSS behavior is therefore expected to emerge far from the axis (for $\beta\geq2$), or even on the axis but far from the head (for $\beta=2$).
Far from the axis (and thus also from the head), the shock boundary conditions on $\chi=1$ asymptote to the GSS form
\begin{equation}
\label{eq:GSS_BCs}
\myQ_s =\frac{1}{2\beta^2} \coma \quad
\myP_s =\frac{2}{3\beta^2} \coma \quad
\mbox{and} \quad
\myU_s =\beta \fin
\end{equation}

Finite energy solutions exist only for $\beta>2$.
Indeed, in \S\ref{sec:NumericalSolution}, we numerically find that the \BoldA{full (}self-similar \BoldA{but} generally non-GSS) solution asymptotes to $\beta\simeq 2.02$ far from the axis.
Before addressing such $\beta>2$ solutions, we first discuss the special case $\beta=2$, for which an analytic flow solution can be found.
Although the GSS equations are in principal valid in this case even as $\eta\to0$, the solution itself is \BoldA{shown} to diverge near the axis.

\subsection{Analytic GSS solution for $\beta=2$}
\label{sec:beta2}

\begin{figure*}
	\centerline{
{\epsfxsize=7.5cm \epsfbox{\myfig{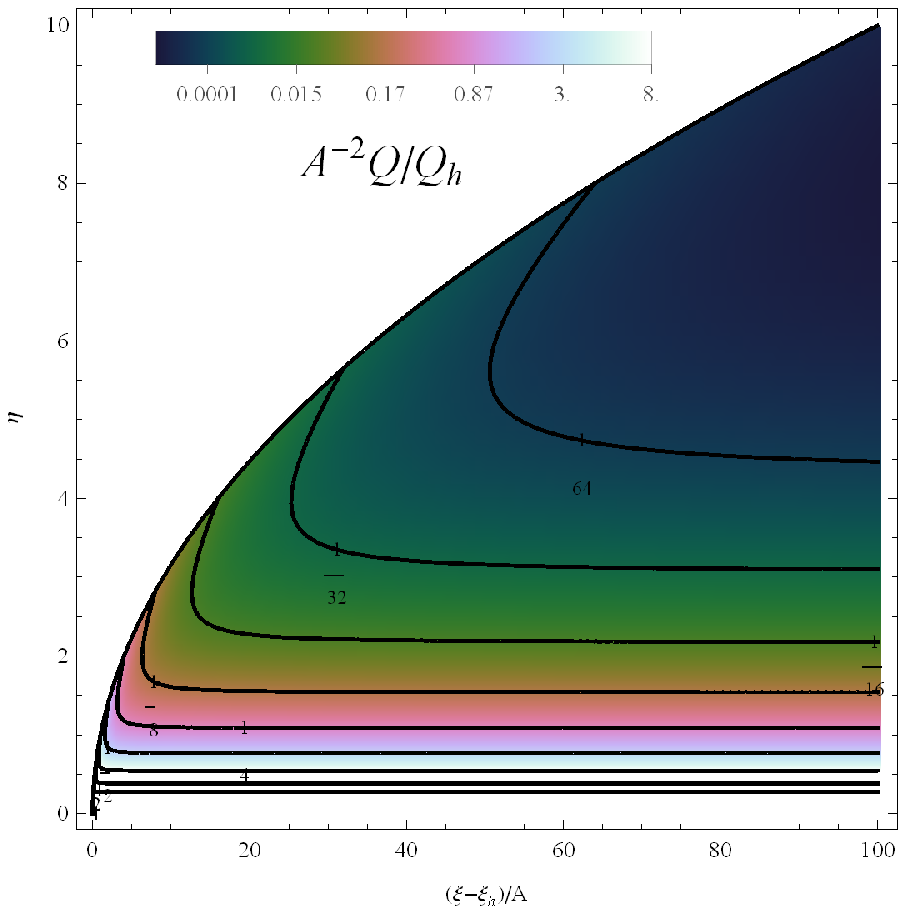}}}
{\epsfxsize=7.5cm \epsfbox{\myfig{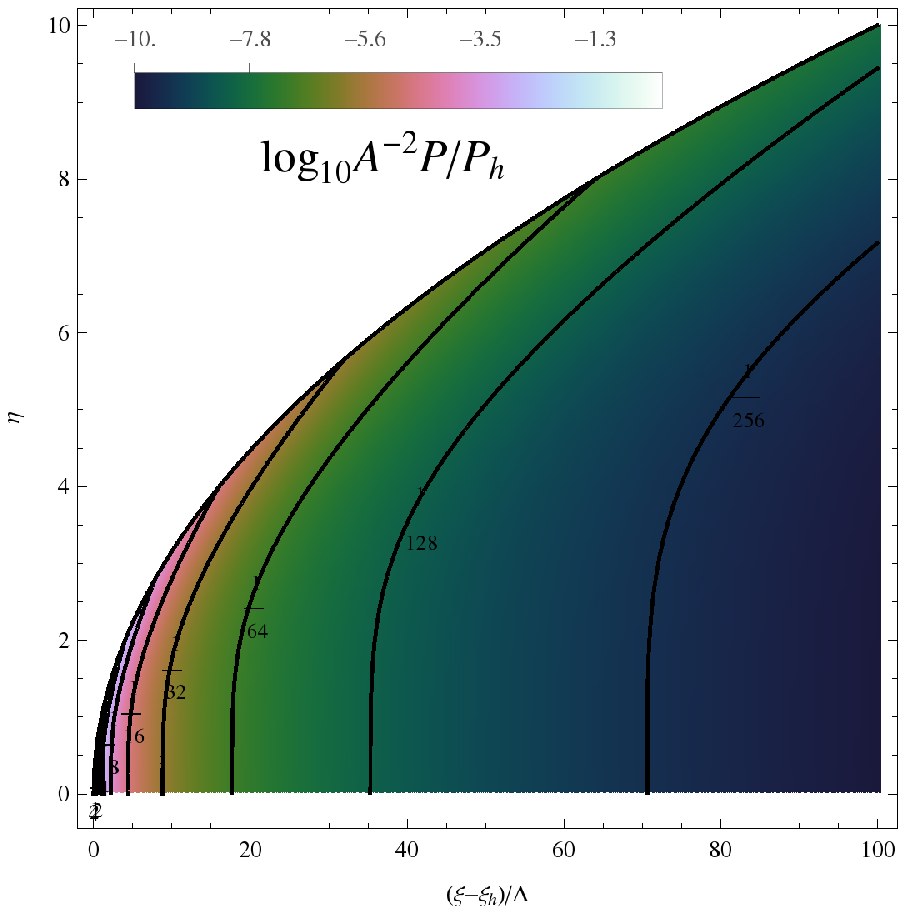}}}
}
	\centerline{
{\epsfxsize=7.5cm \epsfbox{\myfig{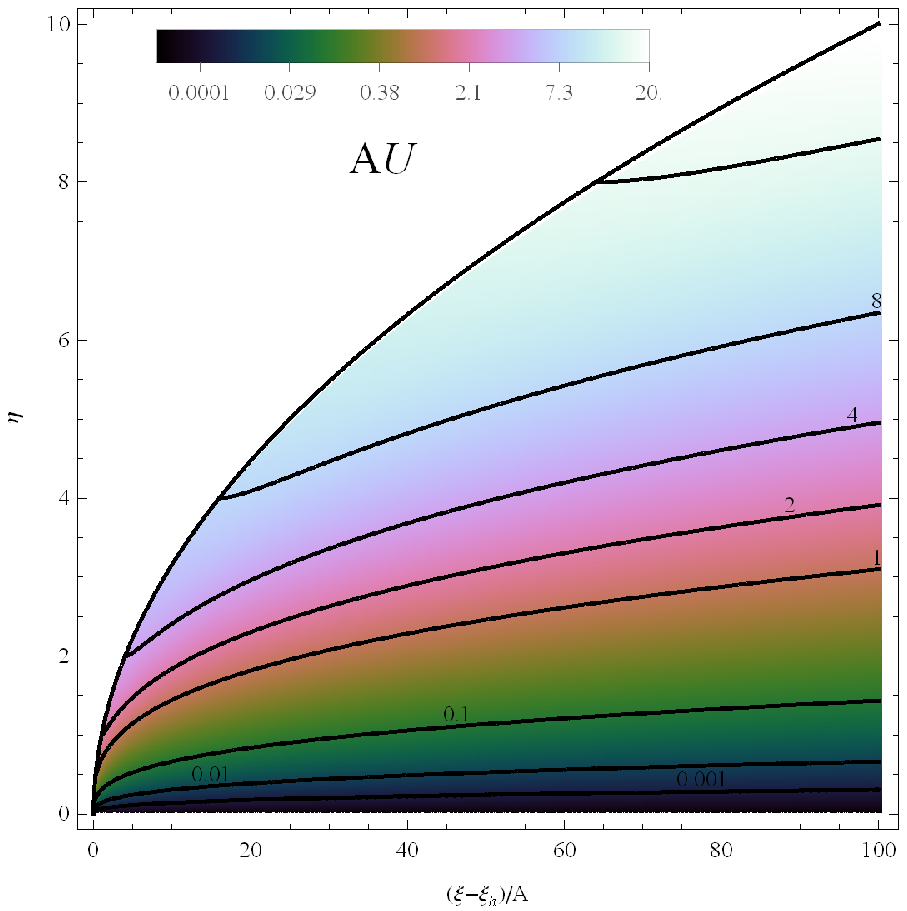}}}
{\epsfxsize=7.5cm \epsfbox{\myfig{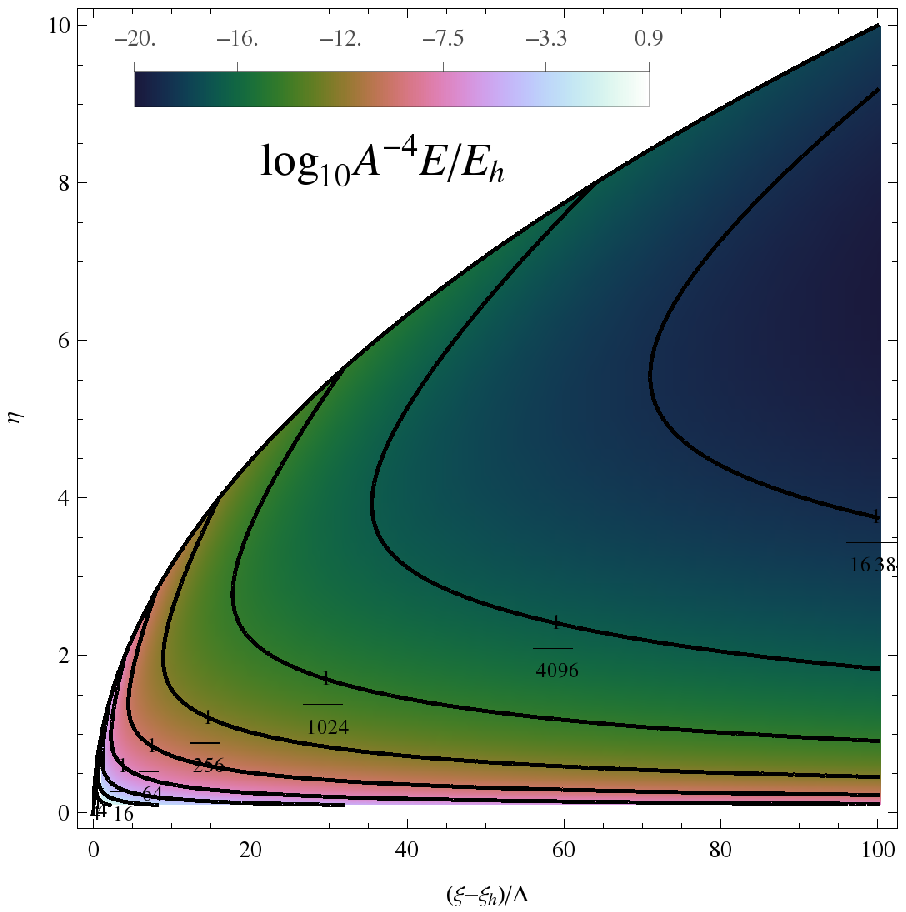}}}
}
\caption{ \label{fig:GSS_beta2}
Self-similar jet in the GSS regime, for the special, analytically solvable case $\beta=2$.
\BoldA{The azimuthal cross section} shows the self-similarly scaled Lorentz factor squared $Q$, proper pressure $P$, radial velocity $U$, and energy density $E$ (see labels). Colormaps \citep[cubehelix;][]{Green11} and contours (in interval factors of 2 for $Q,P$ and $U$, and in factors of 4 for $E$) show each quantity, scaled by the shock width normalization $A$ (see Eq. \ref{eq:SS_parameterB}), with $P,Q$ and $E$ normalized also by their head values.
Note the divergence of $Q$ \BoldA{(and so, also of $E$)} toward the axis.
% JetAnalytic/JetNew121
}
\end{figure*}

\begin{figure*}
	\centerline{
{\epsfxsize=7.5cm \epsfbox{\myfig{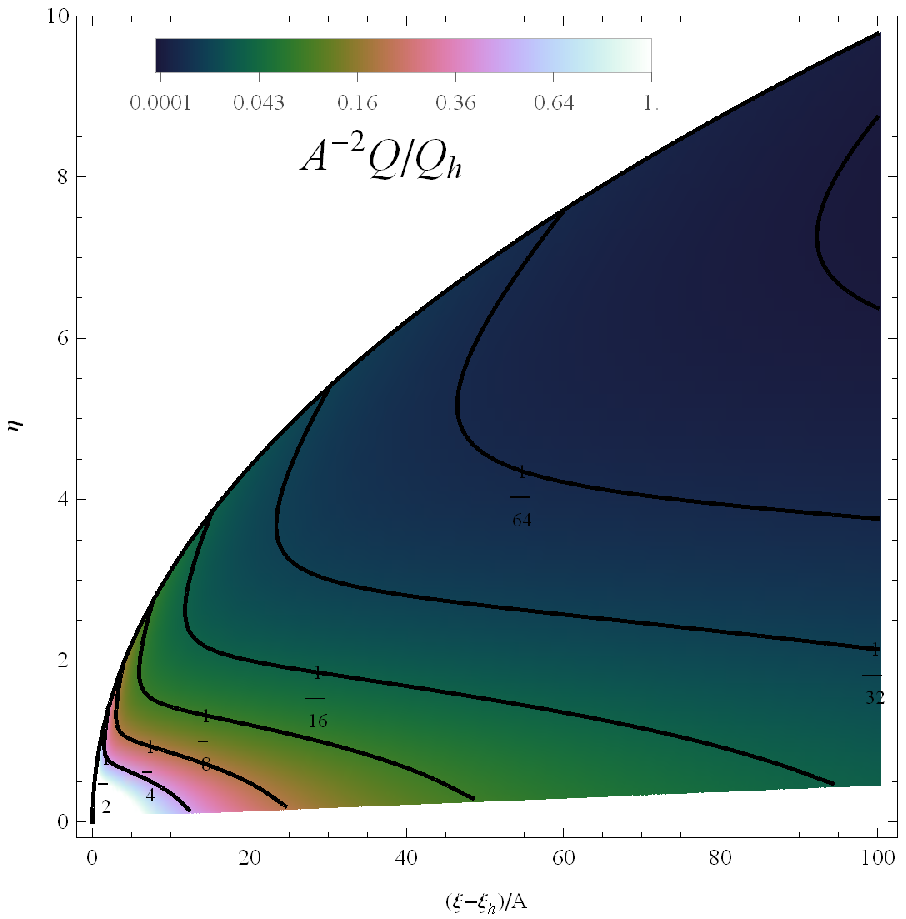}}}
{\epsfxsize=7.5cm \epsfbox{\myfig{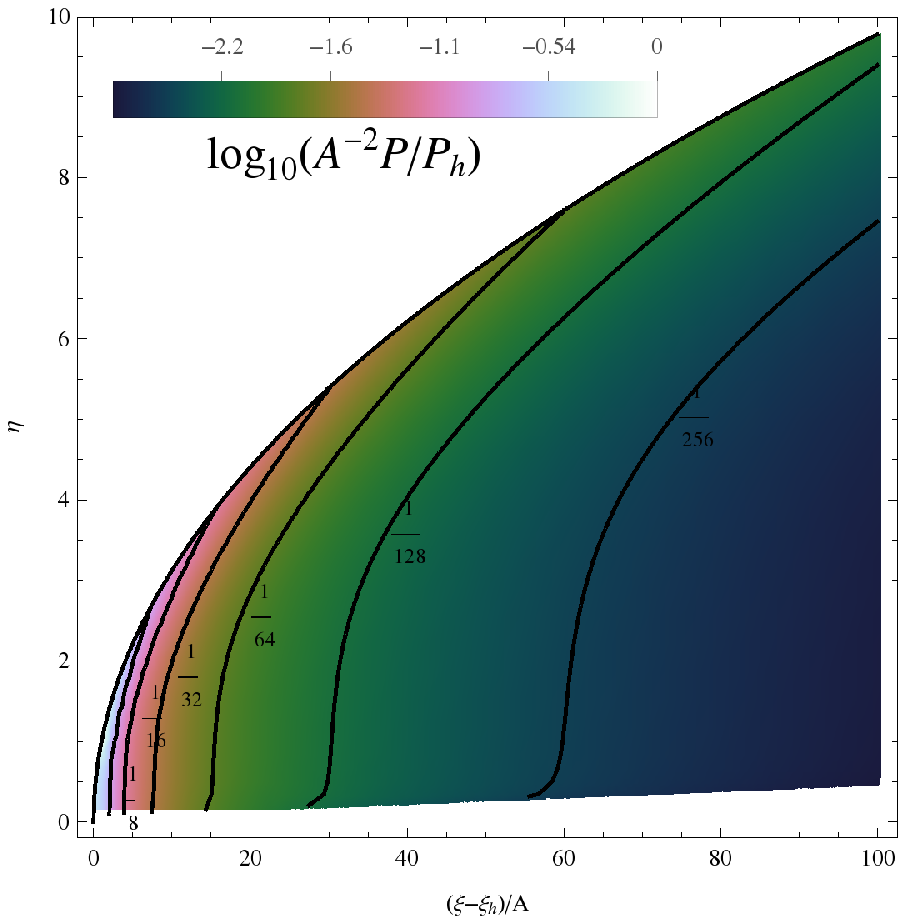}}}
}
	\centerline{
{\epsfxsize=7.5cm \epsfbox{\myfig{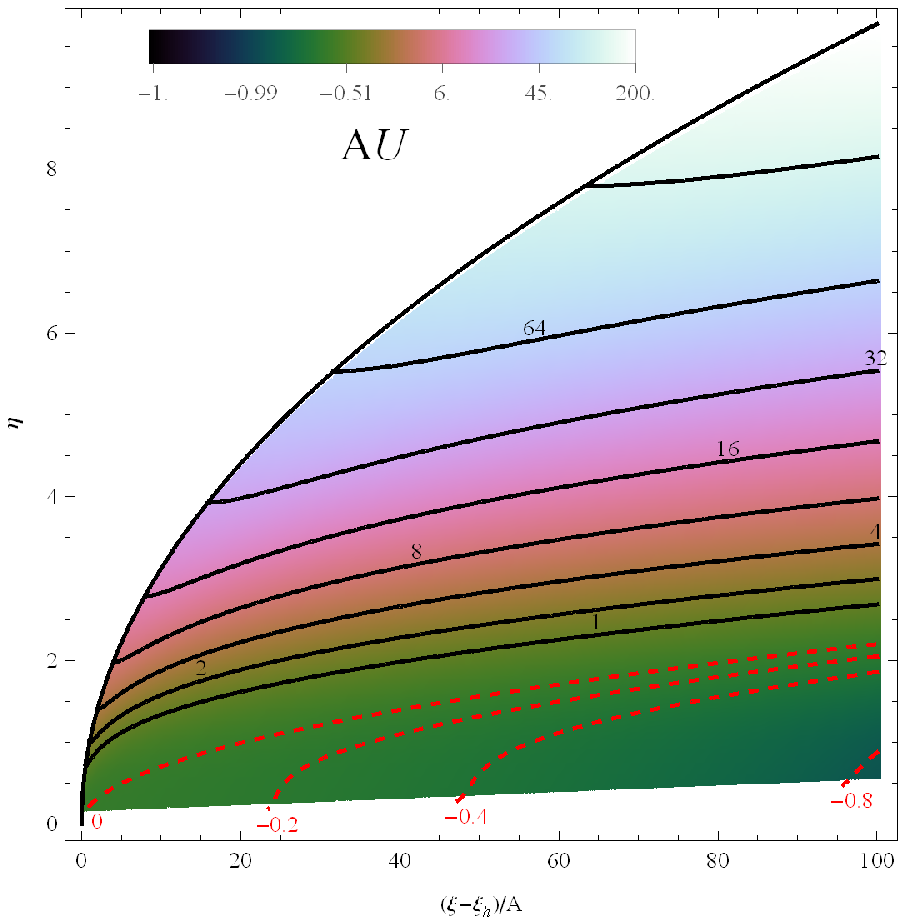}}}
{\epsfxsize=7.5cm \epsfbox{\myfig{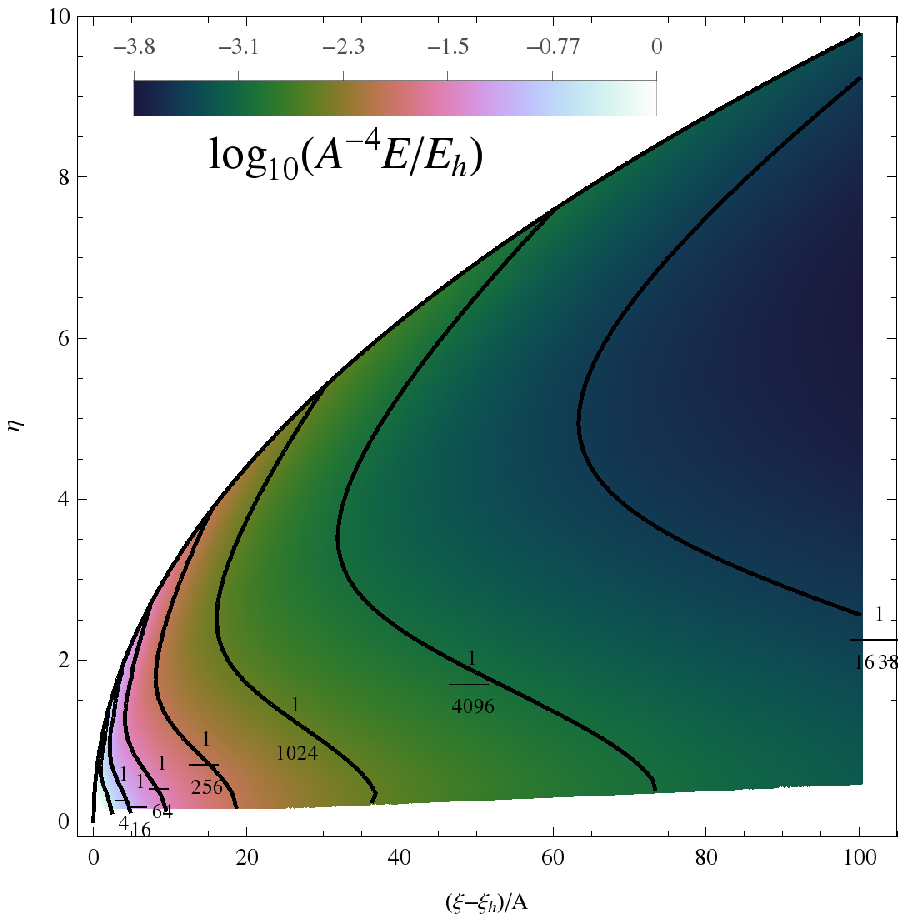}}}
}
\caption{ \label{fig:Beta202}
GSS profiles for the numerically-motivated case $\beta=2.02$. Notations are as in Figure \ref{fig:GSS_beta2}.
Notice the radial inflow ($U<0$) emerging near the axis, characteristic of $\beta>2$.
% JetAnalytic/JetNew133
}
\end{figure*}

\BoldA{As mentioned above,} in the special case $\beta=2$, the last term in each of Eqs.~(\ref{eq:GSS_t})--(\ref{eq:GSS_r}) vanishes.
The equations can then be solved analytically far from the head, where the $\xi_h$ terms are negligible.
For the boundary conditions (\ref{eq:GSS_BCs}), the solution is
\begin{align} \label{eq:beta2approx}
\myU  = & 6\chi-\myfa  \nonumber  -\sqrt{4 g_1^{-1}\chi(4\chi^2+27) +12\chi ^2-54-\myfa ^2} \coma \nonumber \\
\myQ  = & (8\chi \myU -2\myU ^2)^{-1} \coma \\
\myP  = & \frac{1}{6}\exp\left[\int_1^\chi  \left( \frac{4}{\myU -4\bar{\chi}}+\frac{\myU '}{2\bar{\chi}-\myU }
\right)\,d\bar{\chi}\right] \nonumber \coma
\end{align}
where for brevity we defined $\myfa \equiv [4\chi ^2+3(\myfb -3)^2/\myfb ]^{1/2}$ and
$\myfb \equiv[27+32\chi ^4+8\chi ^2(27+16\chi ^4)^{1/2}]^{1/3}$.
\BoldA{This} solution is shown in Figure \ref{fig:GSS_beta2}\BoldA{, as an azimuthal cross section through the jet.}

Far from the shock front ($\chi \gg 1$), the leading terms in this solution are
\begin{eqnarray} \label{eq:SSS_scaling}
\myQ  \simeq \frac{1}{27} + \frac{3}{64\chi^{2}} & \quad \to \quad & Q \simeq
\frac{1}{27\eta^{2}A^2}, \nonumber \\
\myP  \simeq \frac{0.046}{\chi} & \quad \to \quad & P \simeq \frac{0.046}{A\xi}, \nonumber \\
\mbox{and} \quad \myU  \simeq \frac{27}{8\chi} & \quad \to \quad & U \simeq \frac{27\eta^{3}A^2}{8\xi} \fin
\end{eqnarray}
Hence, although Eq.~(\ref{eq:beta2approx}) provides an exact solution to the
GSS equations when $\xi_h\to0$, \BoldA{and so gives an asymptotic solution for $\xi\gg\xi_h$,} this solution is not physical on the axis, where $Q$ diverges.
This local divergence is more severe than, and is not directly related to, the \BoldA{global} logarithmic divergence of the total energy, associated with the \BoldA{marginally} wide ($\beta=2$) shock profile.

\begin{figure*}[t]
	\centerline{
{\epsfxsize=5.5cm \epsfbox{\myfig{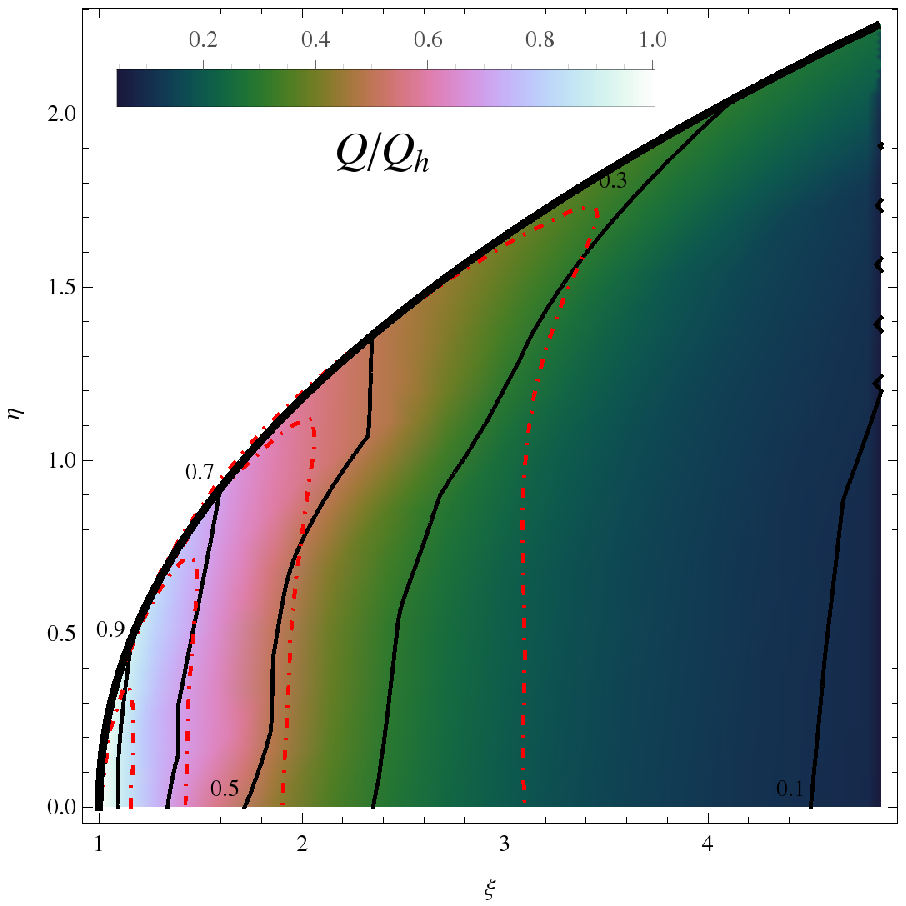}}}
{\epsfxsize=5.5cm \epsfbox{\myfig{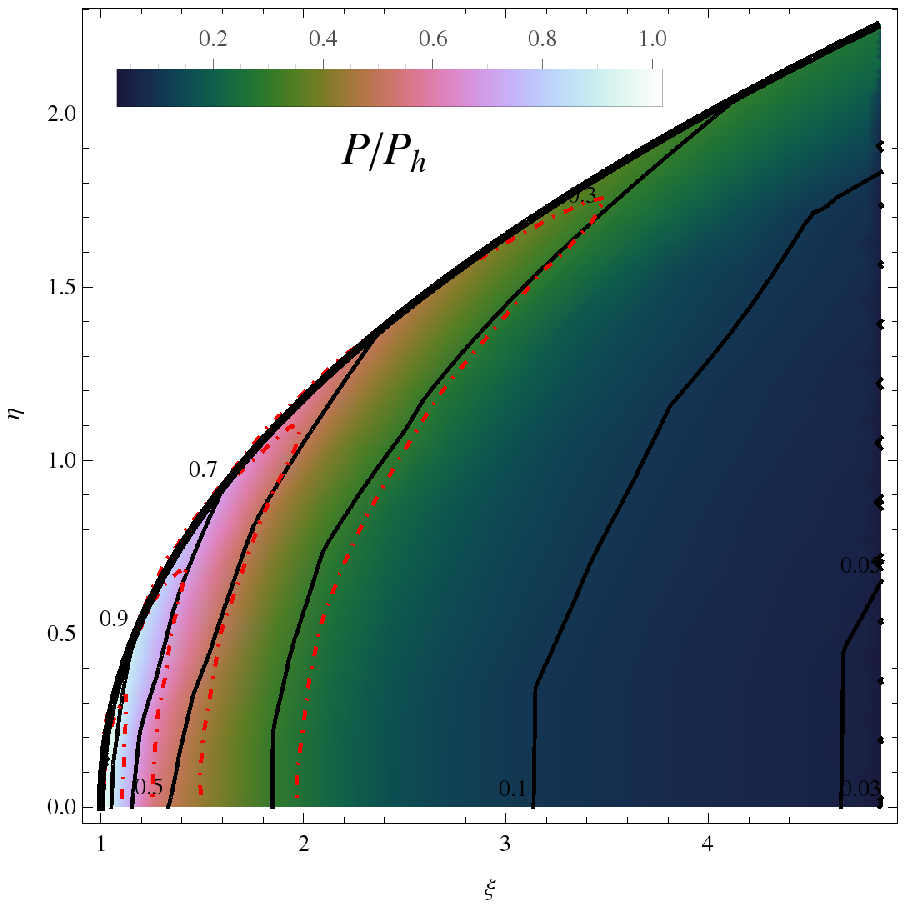}}}
{\epsfxsize=5.5cm \epsfbox{\myfig{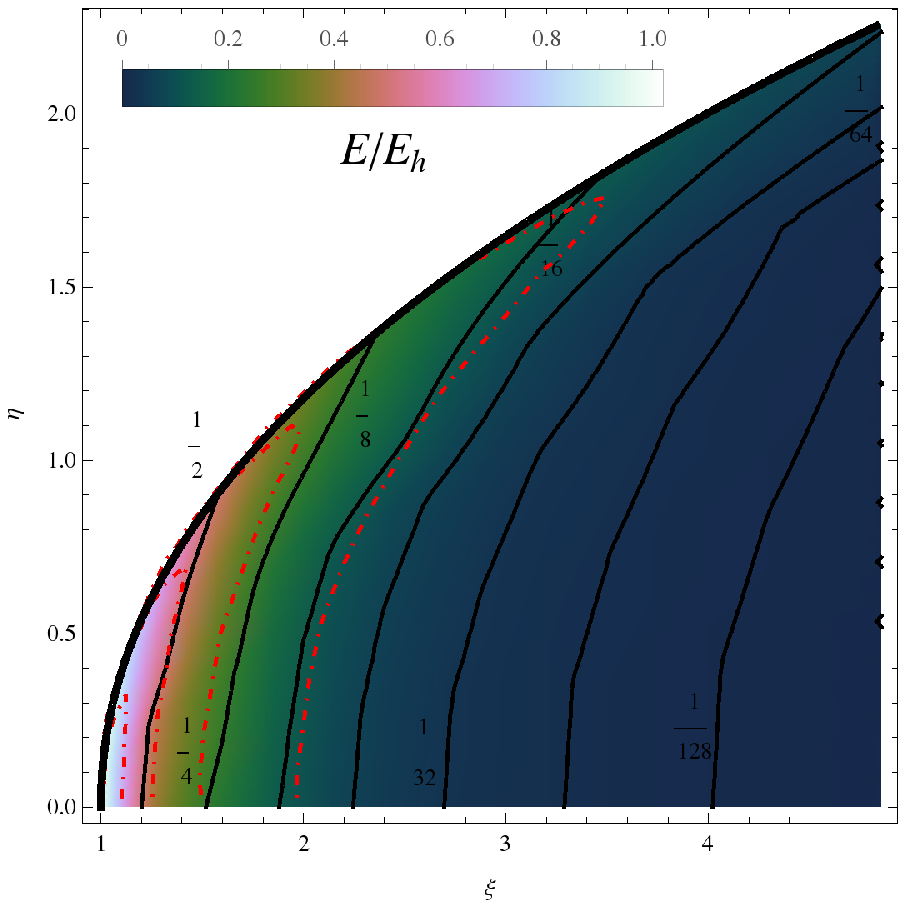}}}
}
\caption{ \label{fig:HeadMono}
Head structure, found by approximately solving the self-similar equations under the monotonicity constraint (see \S\ref{sec:MonotonicHead}).
The profiles are in approximate agreement with the G07 simulations (dot-dashed contours).
Notations are as in Figure \ref{fig:GSS_beta2}. \BoldA{Here,} the shock profile parameters $\beta$ and $A$ are determined self-consistently by the solution.
}
% JetSemi/HeadModel36
\end{figure*}

Nevertheless, the $\beta=2$ solution does provide an adequate analytic approximation of the jet in the outer envelope region, where $\chi$ is not too large.
For example, in a radial cross section, it shows a minimal $Q$ where $0=\pr_\eta Q \propto \beta\chi\myQ'+2(\beta-1)\myQ$.
This occurs at $\chi\simeq 1.61$, or equivalently at a fraction
\begin{equation} \label{eq:QminBeta2}
f\equiv \frac{\eta}{\eta_s} = \chi^{-1/\beta} \simeq 0.79
\end{equation}
of the shock radius, close to the numerical value found \BoldA{for the full solution} in \S\ref{sec:NumericalSolution}.
Moreover, its $P$ and $U$ profiles (see \fig{fig:GSS_beta2}) agree qualitatively with \BoldA{the full} solution even in the head (Figures \ref{fig:HeadMono} and \ref{fig:Head}) region, as well as with the simulated (G07) head.

\subsection{GSS with $\beta=2.02$}
\label{sec:beta202}

As the $\beta\leq 2$ GSS regime is ruled out in \S\ref{sec:beta2} and in \S\ref{sec:NoBetaLess2}, here we consider $\beta$ slightly larger than $2$.
The numerical solution in \S\ref{sec:NumericalSolution} indicates that far from the head, the shock indeed approaches a $\beta\simeq 2.02$ profile.
Accordingly, we now derive the GSS solution for $\beta=2.02$, but point out that the qualitative features of the solution are not sensitive to the precise value of $\beta$ in the $2<\beta<3$ range.

In this range of $\beta$, the GSS Eqs.~(\ref{eq:GSS_t}--\ref{eq:GSS_BCs}) form a closed system of ordinary differential equations, which can only be solved numerically.
The solution for $\beta=2.02$ is shown in Figure \ref{fig:Beta202}.

As the figure shows, while the $P$ profile is not qualitatively changed with respect to the analytic $\beta=2$ case, the profiles of $Q$ and $U$, and subsequently also of $E$, are significantly altered.

Most importantly, $Q$ no longer diverges near the axis, and never becomes a function of $\eta$ alone.
Such \BoldA{$\beta>2$} solutions are therefore physical, unlike the diverging $\beta=2$ profile, and can in principle be matched to the near-axis solution.

Interestingly, the $U$ profile is not everywhere positive, \ie the radial flow is not everywhere an outflow.
Unlike the $\beta=2$ case, $U$ becomes negative at large $\chi$, corresponding to a radial inflow converging on the axis.
For $\beta=2.02$, the transition occurs at $\chi\simeq 20.3$, or equivalently at $f\simeq 0.23$.
The exact location of the transition is sensitive to the precise value of $\beta$.
For example, $\beta=2.04$ gives  $f\simeq 0.32$.

The minimum of $Q$ along a radial cross section is found at $f\simeq 0.77$, not far from the corresponding minimum Eq.~(\ref{eq:QminBeta2}) in the $\beta=2$ case.
This result is less sensitive than the $U=0$ contour to the precise value of $\beta$, giving for example a similar, $f=0.75$ \BoldA{ value} for $\beta=2.04$.

\subsection{Axial expansion and its monotonicity constraint}
\label{sec:Axis}

As shown in \S\ref{sec:beta2} and \S\ref{sec:beta202} above, the axial region of the jet is distinct from the GSS envelope.
It is also distinguishable from the head region, as shown in \S\ref{sec:MonotonicHead} below.
Indeed, the axial structure shows that the monotonic nature of the flow near the head, in which the fields $\myF$ increase monotonically toward the shock in both the $(-\xi)$ and $\eta$ directions, cannot hold far beyond the head region.

To see this, expand $Q$ and $P$ near the axis in even powers of $\eta$,
\begin{equation} \label{eq:Qexpansion}
Q(\xi,\eta)=Q_a(\xi)+Q_2(\xi)\eta^2+Q_4(\xi)\eta^4+\ldots
\end{equation}
and
\begin{equation} \label{eq:Pexpansion}
P(\xi,\eta)=P_a(\xi)+P_2(\xi)\eta^2+P_4(\xi)\eta^4+\ldots \coma
\end{equation}
and expand $U$ in odd powers of $\eta$,
\begin{equation} \label{eq:Uexpansion}
U(\xi,\eta)=U_1(\xi)\eta+U_3(\xi)\eta^3+U_5(\xi)\eta^5+\ldots \fin
\end{equation}
\BoldA{Here, the $F_n$ are numerical factors.}
An $\mathcal{O}(\eta)$ expansion of the flow equations near the axis now shows that
\begin{align}\label{eq:U1res}
U_1(\xi) & = \partial_\eta U(\xi,\eta=0) \\
& =
\frac{Q_{a}'}{Q_{a}^2}-1+\frac{\frac{9}{2}-Q_{a}'\left(4\xi^2+\frac{5}{4Q_{a}^2}\right)} {1+4\xi Q_{a}}
\fin \nonumber
\end{align}

For large $\xi\gg\xi_h$, one can approximate the axial behavior of $Q$ as a power-law, $Q_a\simeq Q_{a0}\xi^{-\alpha}$, \BoldA{where $Q_{a0}$ is a constant,} so Eq.~(\ref{eq:U1res}) gives
\begin{equation}
U_1(\xi\gg \xi_h) \simeq \frac{7}{2}-5\alpha+\frac{\alpha\xi^{\alpha-1}}{4Q_{a0}}+\frac{6(4\alpha-3)Q_{a0}\xi}{4Q_{a0}\xi+\xi^{\alpha}} \fin
\end{equation}
If $\alpha>1$, then $U_1(\xi\gg \xi_h)$ diverges due to the third term.
In addition to being nonphysical, this also breaks monotonicity, as at the head $U_1$
% $U_1\simeq 2A$
is finite. 	
If $\alpha<1$, then $U_1(\xi\to\infty)=\alpha-1<0$. This rules out monotonicity as well, and is consistent with the $U<0$ inflow inferred near the axis \BoldA{from the GSS analysis} (see \S\ref{sec:beta202}).

A fully monotonic flow near the axis is thus possible only \BoldA{for the special case} $\alpha=1$, such that
\begin{equation}
U_{1}=\frac{1-2Q_{a0}}{4Q_{a0}+16Q_{a0}^{2}} = \const \geq 0 \coma
\end{equation}
which \BoldA{also} requires $Q_{a0}\leq1/2$.
We conclude that only a $Q_a$ profile with $\alpha=1$ and $Q_{a0}\leq 1/2$ can yield a fully monotonic behavior near the axis.
However, as mentioned in \S\ref{sec:AnalyticGeneral}, such  monotonicity cannot persist radially out to the shock, as this would contradict the faster decline of $Q_s(\xi)$ according to the $\beta>2$ shock boundary conditions.

\subsection{Monotonic head region}
\label{sec:MonotonicHead}

Near the head, previous simulations (G07) and our numerical solution (\S\ref{sec:NumericalSolution}) suggest a monotonic behavior. This has several interesting implications.

Near the head, the shock profile is \BoldA{nearly} parabolic, so one may approximate $\xi_s\simeq \xi_h+A\eta^2$.
\BoldA{Here, the arbitrary constant $A$ coincides with that defined in Eq.~(\ref{eq:SS_parameterB}), for $\beta\to 2$.}
Equations (\ref{eq:SSt},\ref{eq:SSBC},\ref{eq:ShockDerivative}) then yield
\begin{equation}
\partial_\eta Q\simeq \frac{A(2-3A)}{4}\eta
\quad \mbox{and} \quad
\partial_\xi Q \simeq \frac{A-1}{4} \coma
\end{equation}
so imposing monotonicity would imply that the jet cannot be too narrow, \ie that $A<2/3$.
The jet cannot be too wide, either; $A\lesssim 0.1$ ($A\lesssim 0.3$) can be shown semi-analytically (numerically) to lead $Q$ to vanish near the head; see \S\ref{sec:WideJets}.

One may attempt to solve the self similar equations for a monotonic flow.
To do so, we numerically minimize the sum of the squares of the left hand sides of Eqs.~(\ref{eq:SSt}--\ref{eq:SSr}), while imposing the boundary conditions \ref{eq:SSBC}, and, in addition, constraining the fields to be monotonic.
This leads to an approximate solution, illustrated in Figure \ref{fig:HeadMono}.

The resulting monotonic profiles qualitatively resemble the structure of the head found numerically and in G07, as shown in the figure.
We find such solutions only for $\xi\lesssim 5$, beyond which the monotonic fields diverge.
This suggests that $\xi_g\simeq 5$ roughly marks the edge of the monotonic head region.

An approximate, monotonic description \BoldA{of the flow in the head region} may be found using the axial expansion in Eqs.~(\ref{eq:Qexpansion}--\ref{eq:Uexpansion}); see \S\ref{sec:HeadModel}.
The resulting approximation is plotted on top of the numerical solution of the head region in Figure \ref{fig:Head}, using an approximate shock profile $\xi_s(\eta)$ inferred from the numerical solution \BoldA{of \S\ref{sec:NumericalSolution}}.
As the figure shows, the expansion fits the results rather well near the head.

\section{Numerical solution}
\label{sec:NumericalSolution}

In order to solve the self-similar flow Eqs.~(\ref{eq:SSt}--\ref{eq:SSBC}) numerically, we expand the shock profile $\eta_s(\xi)$ to increasingly high order, and find the optimal jet solution at each order, in two different methods.
Following the arguments of \S\ref{sec:Analytic}, an optimal solution is defined as the most regular solution to the flow equations, \BoldA{\ie the solution} diverging farthest either from the head or from the shock.
The results of this procedure indicate that a unique solution, which remains regular infinitely far from the head and from the shock, indeed exists.

\subsection{Method}
\label{sec:NumericalMethod}

\begin{figure*}
	\centerline{
{\epsfxsize=7.5cm \epsfbox{\myfig{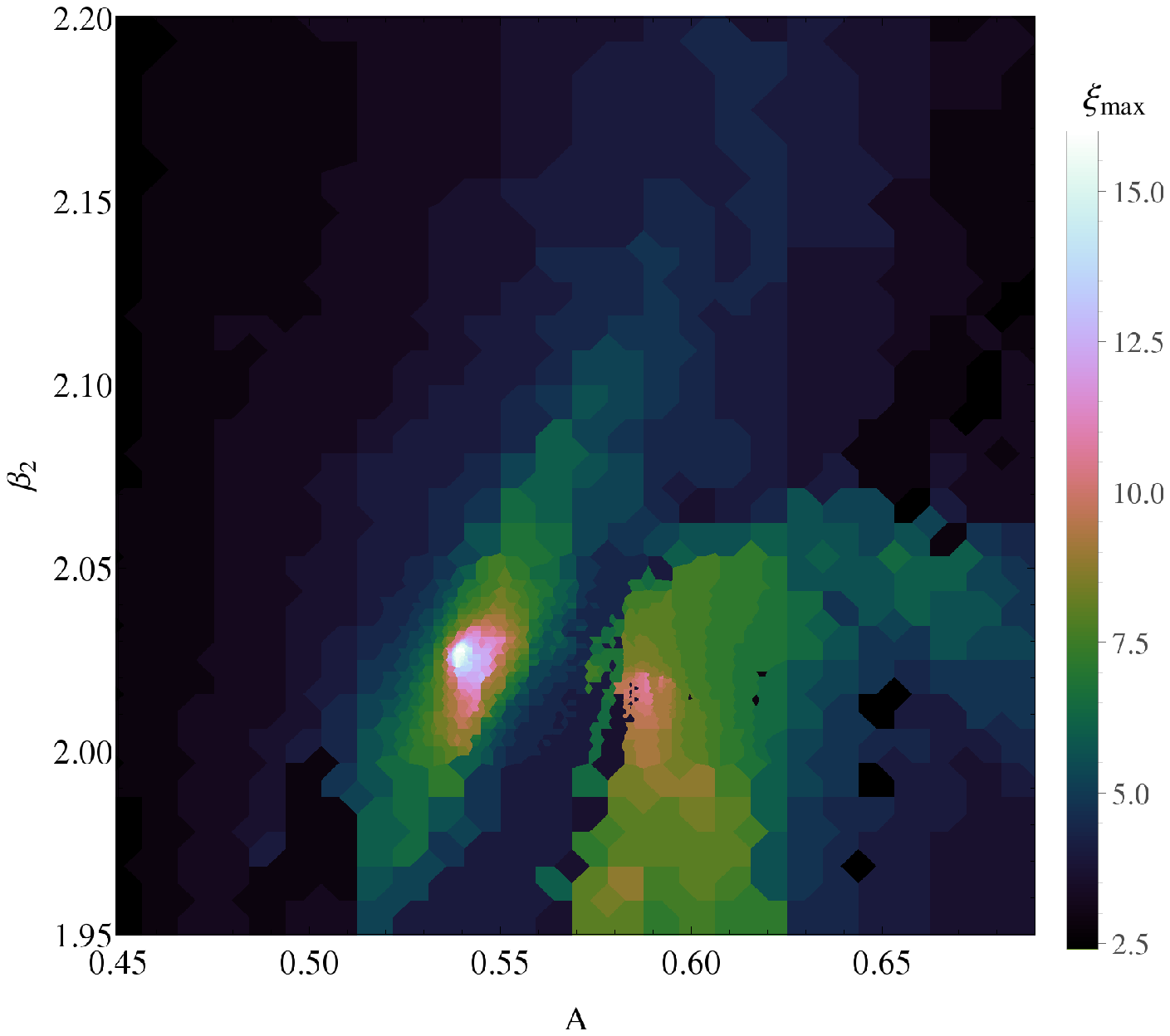}}
% JetSemi/NDSolveXiMaxScan48
}
{\epsfxsize=7.5cm \epsfbox{\myfig{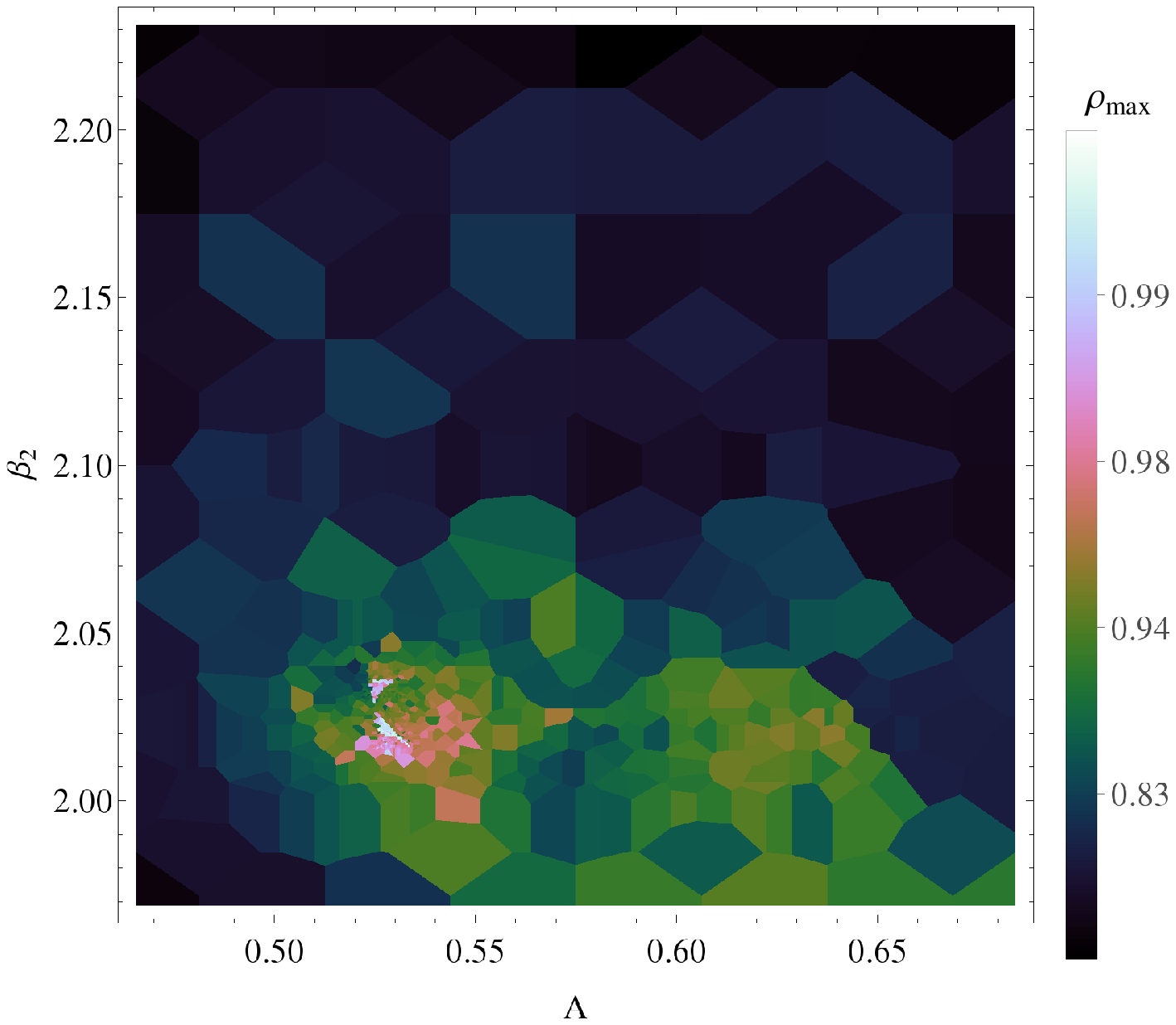}}
% JetSemi/NDSolveRhoMaxScan15
}
}
\caption{ \label{fig:ParameterScan}
Independent scans of $\xi_{max}$ (method 1; left) and $\rho_{max}$ (method 2; right) in the $A$--$\beta_2$ phase space, for a second-order shock profile $\xi=(1+A\eta_s^2)^{1+(-1+\beta_2/2)\mysigma^2}$ (see \S\ref{sec:NumericalMethod} for details).
We find the maximal $\xi_{max}$ for $\{A\simeq0.54, \beta_2\simeq2.027\}$,
and the maximal $\rho_{max}$ for $\{A\simeq0.53, \beta_2\simeq2.020\}$.
}
% JetSemi/HeadModel36
\end{figure*}

First, we map the (azimuthal cross section of the) jet onto the unit square, $0\leq\{\rho,\sigma\}\leq1$, through the transformation
\begin{equation}
\rho\equiv 1-\frac{\eta}{\eta_s} \quad \mbox{and} \quad \mysigma\equiv 1-\frac{\xi_h}{\xi} \fin
\end{equation}
This maps the shock and the axis respectively onto $\rho=0$ and $\rho=1$.
The head and infinite downstream ($\xi\to\infty$) are similarly mapped onto $\mysigma=0$ and $\mysigma=1$.

Next, the shock profile is parameterized as
\begin{equation} \label{eq:ShockProfileAnalytic}
\xi_s = (\xi_h+A\eta^2)^{\beta(\mysigma)/2} \coma
\end{equation}
where we choose this functional form, using $\eta^2$ instead of $\eta$ in the parenthesis, in order to obtain better behaved functions.
We expand the a-priori unknown function $\beta(\mysigma)$ to order $n$,
\BoldA{and switch from a $\xi_s(\eta)$ parametrization to the more general, $\eta_s(\xi)$ description of the shock,}
such that
\begin{equation} \label{eq:ShockProfileExpanded}
\xi = (1+A\eta_s^2)^{1+\epsilon_0+\epsilon_1\mysigma^1+\epsilon_2\mysigma^2+
\ldots + \epsilon_n\mysigma^n } \fin
\end{equation}

Consider the solution for a given order $n$.
The shock profile is defined by the $(n+2)$ \BoldA{undetermined} parameters $A$ and $\epsilon_0,\epsilon_1,\ldots ,\epsilon_n$.
Given some choice of these parameters, one can integrate the equations in two different methods:
\begin{enumerate}
\item
Start from the ($\sigma=0$) head boundary conditions, and advance toward ($\sigma=1)$ downstream infinity.
\item
Start from the ($\rho=0$) shock boundary conditions, and advance toward the ($\rho=1$) axis.
\end{enumerate}

For finite $n$, both methods eventually fail, as the fields diverge at some finite $\mysigma_{max}$ (or equivalently $\xi_{max}$) in method 1, and at some finite $\rho_{max}$ (or equivalently $\eta_{min}$) in method 2.
Indeed, the shock profile must be fine-tuned in order to delay the divergence, and uncover a larger fraction of the jet.

We use both methods, independently maximizing $\xi_{max}$ and $\rho_{max}$ at each order $n$ by scanning the $(n+2)$ dimensional phase space of the shock profile parameters.
Thus, we identify the best approximation to the shock profile at every order.

The results of an order $n=2$ scan are presented in Figure \ref{fig:ParameterScan}, for the maximization of both $\xi_{max}$ (left panel) and $\rho_{max}$ (right panel).
In order to project our four-dimensional scan onto a two-dimensional figure, \BoldA{here} we set $\epsilon_0=\epsilon_1=0$, and vary only $A$ and $\epsilon_2$. It is useful to define \BoldA{the parameter $\beta_2$ through} $\epsilon_2=(-1+\beta_2/2)$; this corresponds to a paraboloid, $\xi_s\simeq \xi_h+A\eta^2$ profile for $\xi\to\xi_h$, smoothly transitioning into a GSS-like, $\xi_s\simeq (\xi_h+A\eta^2)^{\beta_2/2}\sim \eta^{\beta_2}$ profile as $\xi\to\infty$.

Both methods show that the parameters $A\simeq (0.53$--$0.54)$ and $\beta_2\simeq (2.02$--$2.03)$ provide the most accurate shock profile at this order, in the sense that the divergence of the flow takes place farthest from the head or from the shock.
The full $(n+2)$ dimensional optimization process converges with $n$ on a unique shock profile, \BoldA{with similar $A$ and $\beta_2$ values,} as discussed next.

\subsection{Convergence}
\label{sec:NumericalConvergence}

Method number 1, in which $\xi_{max}$ is maximized, is much simpler and faster computationally than the $\rho_{max}$ maximization of method 2.
Therefore, after demonstrating that both methods give similar results at low orders, we pursue high orders using \BoldA{only} the $\xi_{max}$ maximization method 1.
Table \ref{tab:shock_profiles} summarizes the results of this method, providing the parameters of the optimal shock profile at each order $n$.

\begin{table*}
\begin{center}
\begin{tabular}{|l|l|l|l|l|l|l|l|l|l|l|l|l|}
\hline
$n$ & $\xi_{max}$ & $A$ & $\epsilon_0$ & $\epsilon_2$ & \
$\epsilon_4$ & $\epsilon_6$ & $\epsilon_8$ & $\epsilon_{10}$ & \
$\epsilon_{12}$ & $\epsilon_{14}$ & $\epsilon_{16}$ & $\beta_\infty$ \\
\hline
0&12.14&0.5237&0.0323&&&&&&&&&2.0647\\
2&15.93&0.5398&0.0001&0.014&&&&&&&&2.0282\\
4&18.84&0.5489&-0.0047&0.0155&0.002&&&&&&&2.0255\\
6&$>$20.11&0.5433&0.0006&0.0113&0.0036&-0.0015&&&&&&2.0279\\
8&$>$22.17&0.5447&-0.0004&0.0118&0.0019&0.001\BoldA{0}&-0.0018&&&&&2.0253\\
10&$>$24.79&0.5457&-0.0001&0.0107&0.0016&0.0012&-0.0011&-0.0012&&&&2.0221\\
12&$>$24.80&0.5452&-0.0001&0.011\BoldA{0}&0.0015&0.0011&-0.001\BoldA{0}&-0.0011&-0.0001&&&2.0226\\
14&$>$24.98&0.5454&-0.0002&0.0109&0.0016&0.0011&-0.0012&-0.001\BoldA{0}&0.&-0.0001&&2.0222\\
16&$>$25.28&0.5454&-0.0001&0.0109&0.0016&0.0012&-0.0012&-0.001\BoldA{0}&0.&-0.0001&-0.0001&2.0225\\
\hline
\end{tabular}
\end{center}
\caption{\label{tab:shock_profiles}
Optimized shock profile \BoldA{(see Eq.~\ref{eq:ShockProfileExpanded})} for increasing expansion orders $n$.\\}
\end{table*}

The odd $n$ terms do not significantly modify the solution or increase the values of $\xi_{max}$.
We therefore set these terms to zero, and use only the even $n$ terms.
This is equivalent to taking $\beta(\mysigma^2)$ instead of $\beta(\mysigma)$ in \BoldA{the shock parametrization} Eq.~(\ref{eq:ShockProfileAnalytic}).

For convergence tests, we thus define $N\equiv 1+n/2$ as the effective order of the shock expansion.
Our highest order, $n=16$, corresponds to $N=9$, and thus involves searching for the maximum of $\xi_{max}$ in a 10 (including $A$) dimensional parameter space.
Due to the high dimensionality, for orders $n=6$ and above, our maximal $\xi_{max}$ values should be considered as lower limits, as noted in the table.

\begin{figure}
	\centerline{
{\epsfxsize=9cm \epsfbox{\myfig{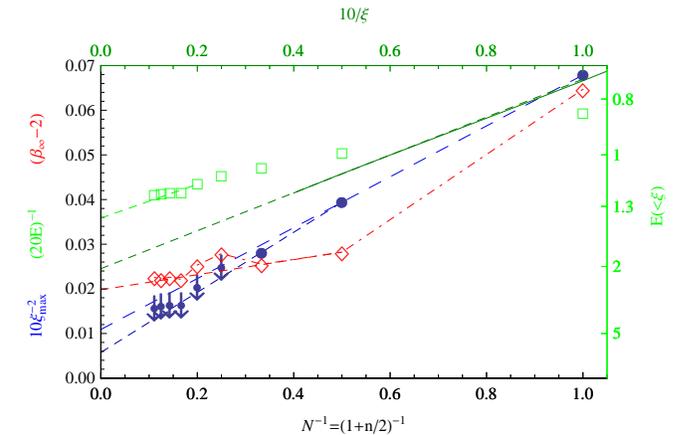}}}
}
\caption{ \label{fig:Convergence1}
Convergence plot (with left and bottom axes) of the shock profiles $\eta_s$ in Table \ref{tab:shock_profiles}, as a function of the effective inverse order $N^{-1}$. As $N$ (or equivalently $n$) increases, $\xi_{max}$ (blue disks; arrows designate lower limits) increases and approaches our resolution limit, $\beta_\infty$ (red diamonds) converges at $\sim2.02$, and $E_{tot}(<\xi_{max})$ (green squares, shown also with the right axis) exceeds $\sim1.4$.
A better estimate of $E_{tot}$ is obtained by plotting $E_{tot}(<\xi)$ for a high ($n=10$) order expansion (solid curve, plotted with upper axis), indicating that $E_{tot}\simeq 2.1$.
These asymptotic estimates are based on extrapolations (dashed curves) to $n\to\infty$ (lower axis) or $\xi\to\infty$ (upper axis).
% JetSemi/NDSolveXiMaxScan55
}
\end{figure}

Figure \ref{fig:Convergence1} shows a convergence plot of our results with the inverse effective order $N^{-1}$ of the expansion.
As the table and figure show, the maximal $\xi_{max}$ (disks in the figure) monotonically increases with $n$, and suggests convergence near our resolution limit (anticipated at $\xi_{max}\sim50$).
The shock profile is well behaved, in the sense that low order $\epsilon_n$ terms do not change significantly as higher order terms are added.

As the order $n$ increases, $\epsilon_0\to0$, corresponding to the expected paraboloid head.
The largest coefficient in the high order $\beta(\sigma)$ expansion is $\epsilon_2\simeq 0.011$, corresponding to $\beta_2\simeq 2.022$. This dominates the deviation from a parabolic profile, and \BoldA{was} therefore chosen as the term scanned in Figure \ref{fig:ParameterScan}.

Far downstream, as $\sigma\to 1$, the shock profile (\ref{eq:ShockProfileExpanded}) approaches a power-law, $\xi_s\propto \eta^{\beta_\infty}$, suggesting a GSS scaling with
\begin{equation}
\beta=\beta_\infty \equiv 2+\sum_{j=0}^n\epsilon_j \fin
\end{equation}
The values of $\beta_\infty$ found at different orders $n$ are shown in the table and in Figure \ref{fig:Convergence1} (diamonds).
These results suggest that the shock profile converges, as $n\to\infty$, at $\beta_\infty\to 2.02$.

The energy $E(<\xi_{max})$ is shown in Figure \ref{fig:Convergence1} (squares plotted against left and bottom axes) to converge slowly.
The figure also shows (solid curve with right and upper axes) a better method to estimate the total energy of the jet, by extrapolating \BoldA{to $\xi\to\infty$} the $E(<\xi)$ profile of a solution with a fixed high $n$ .
The extrapolated result of this \BoldA{(}good\BoldA{)} fit is $E_{tot}\simeq 2.1$; most ($\sim 80\%$) of this energy lies well beyond the head region.

\subsection{Solution existence and uniqueness}
\label{sec:NumericalUniqueness}

\begin{figure*}
	\centerline{
{\epsfxsize=7.5cm \epsfbox{\myfig{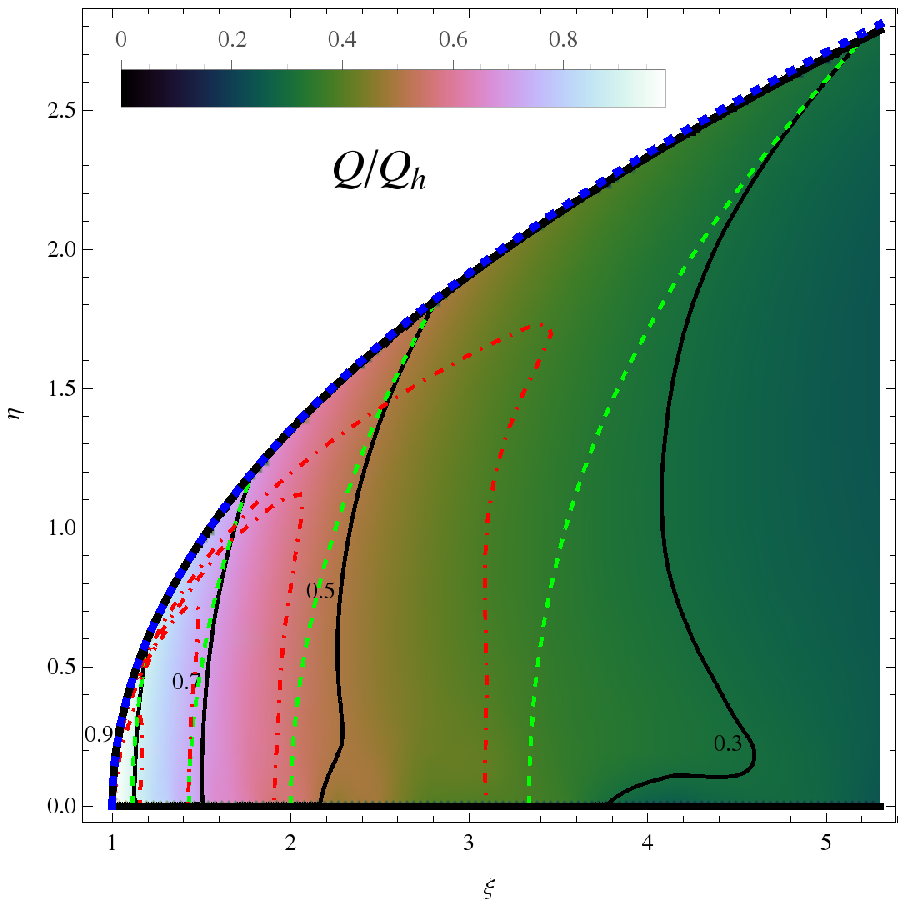}}}
{\epsfxsize=7.5cm \epsfbox{\myfig{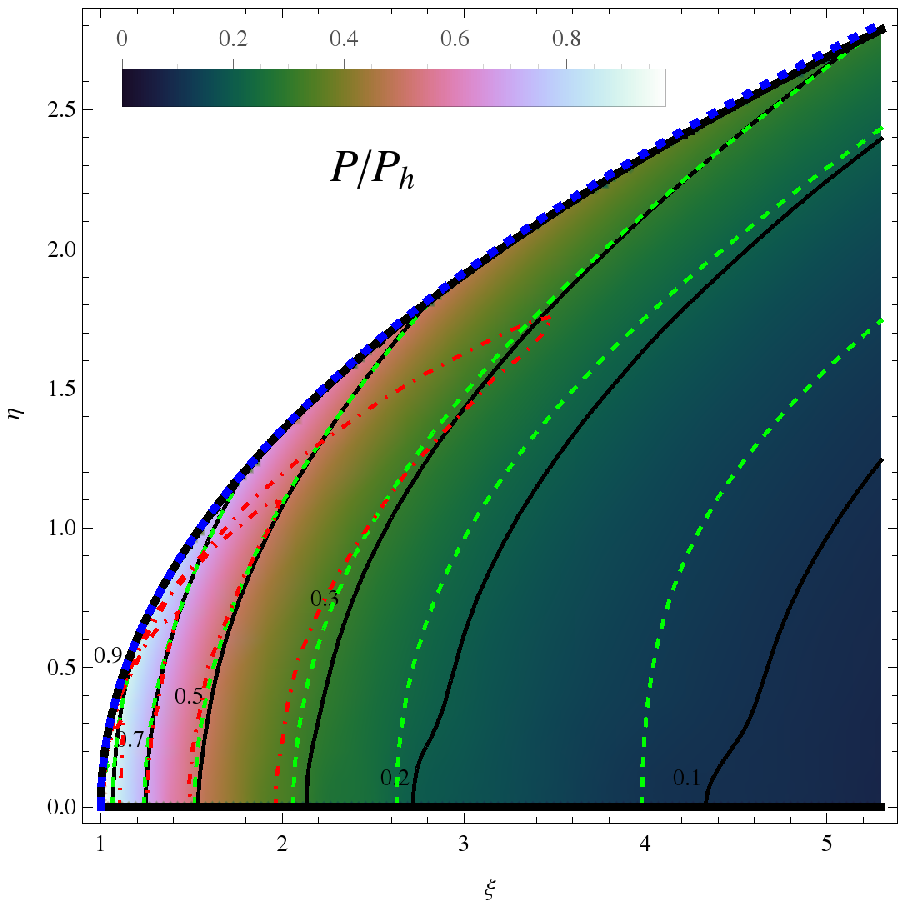}}}
}
	\centerline{
{\epsfxsize=7.5cm \epsfbox{\myfig{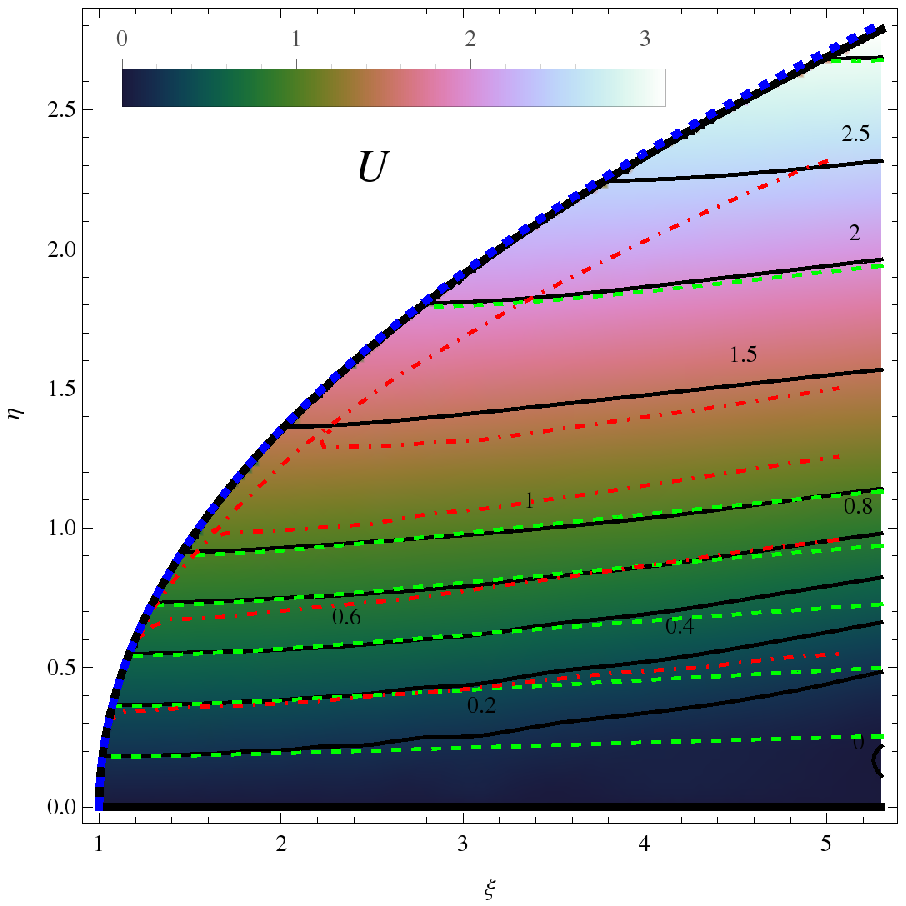}}}
{\epsfxsize=7.5cm \epsfbox{\myfig{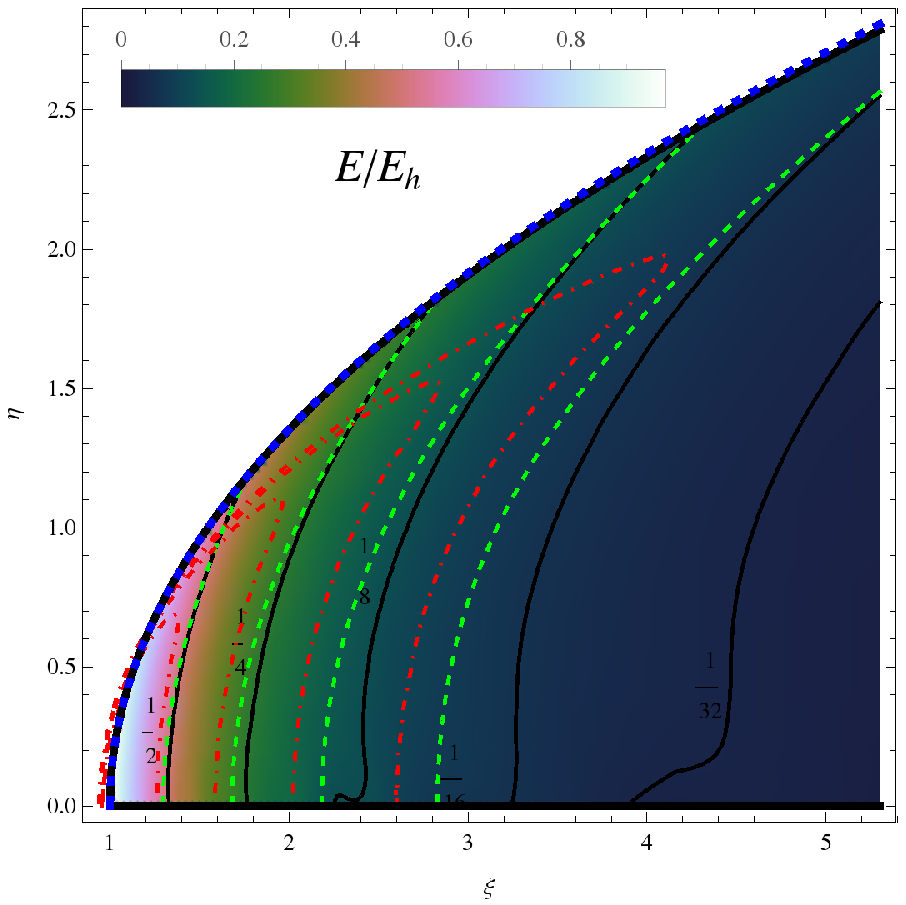}}}
}
\caption{ \label{fig:Head}
Head region of the numerical \BoldA{jet} solution for a high ($n=10$) order expansion of the shock profile, as listed in Table \ref{tab:shock_profiles}.
The analytic model (dashed contours\BoldA{; see \S\ref{sec:HeadModel}}) for the corresponding power-law shock profile (\BoldA{$\beta=2.02$} and $A=0.53$) provides a good fit near the head.
The profiles qualitatively agree with the simulations of G07 (dot-dashed contours) near the head, although the G07 jet is narrower.
Notations are as in Figure \ref{fig:GSS_beta2}, \BoldA{but here} contour values are mostly chosen to match those of G07, for comparison purposes.
}
% JetSemi/HeadModel36
\end{figure*}

\begin{figure*}
	\centerline{
{\epsfxsize=7.5cm \epsfbox{\myfig{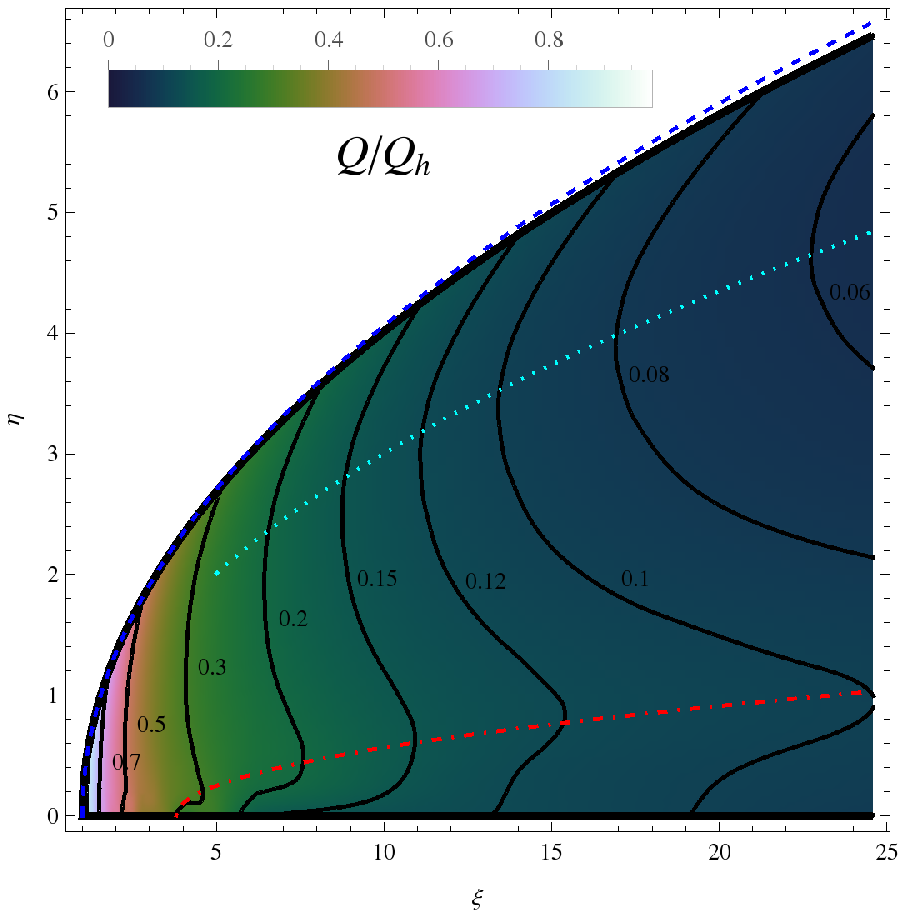}}}
{\epsfxsize=7.5cm \epsfbox{\myfig{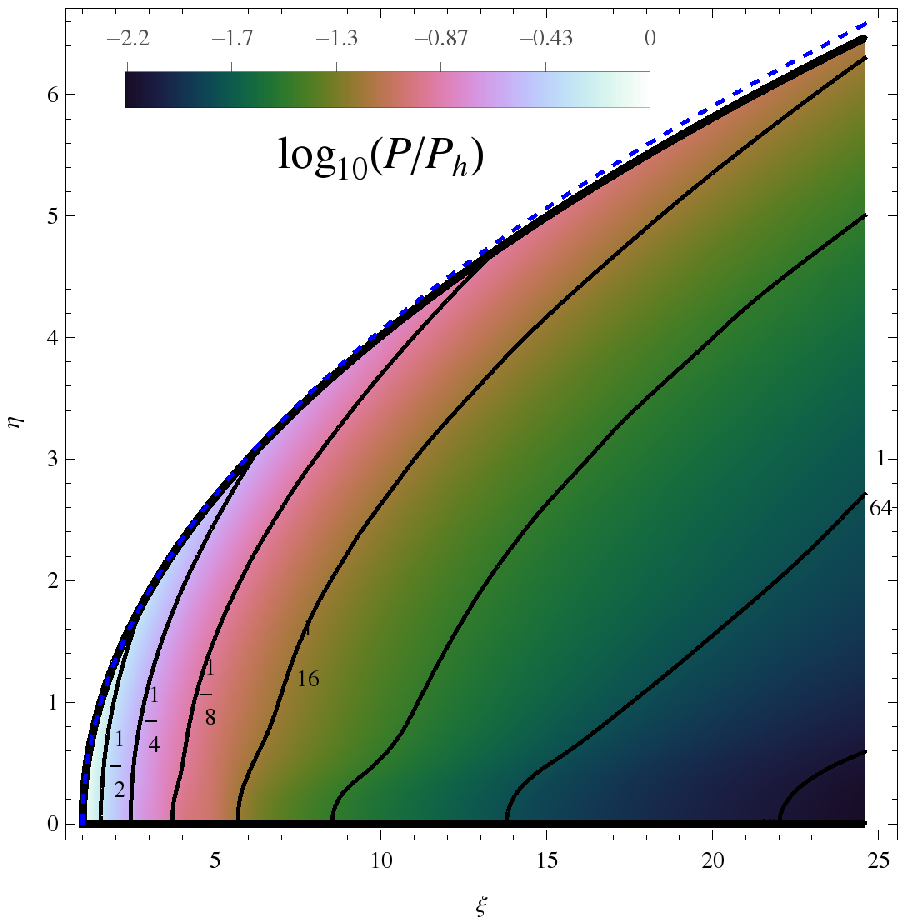}}}
}
	\centerline{
{\epsfxsize=7.5cm \epsfbox{\myfig{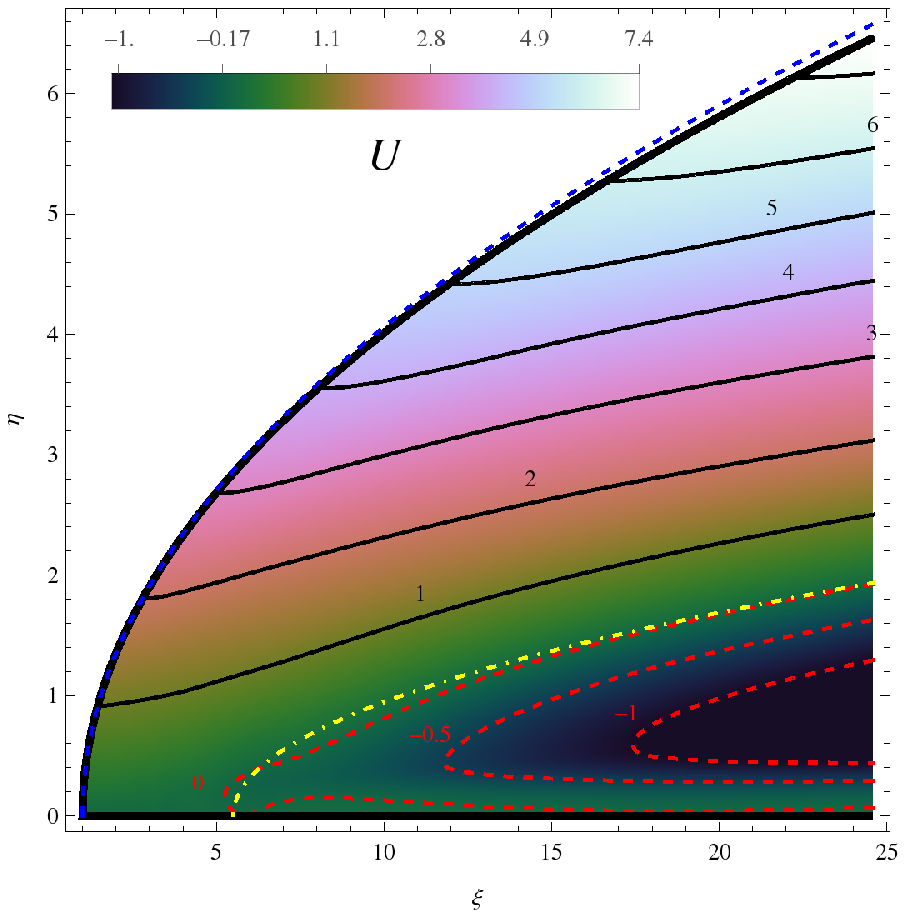}}}
{\epsfxsize=7.5cm \epsfbox{\myfig{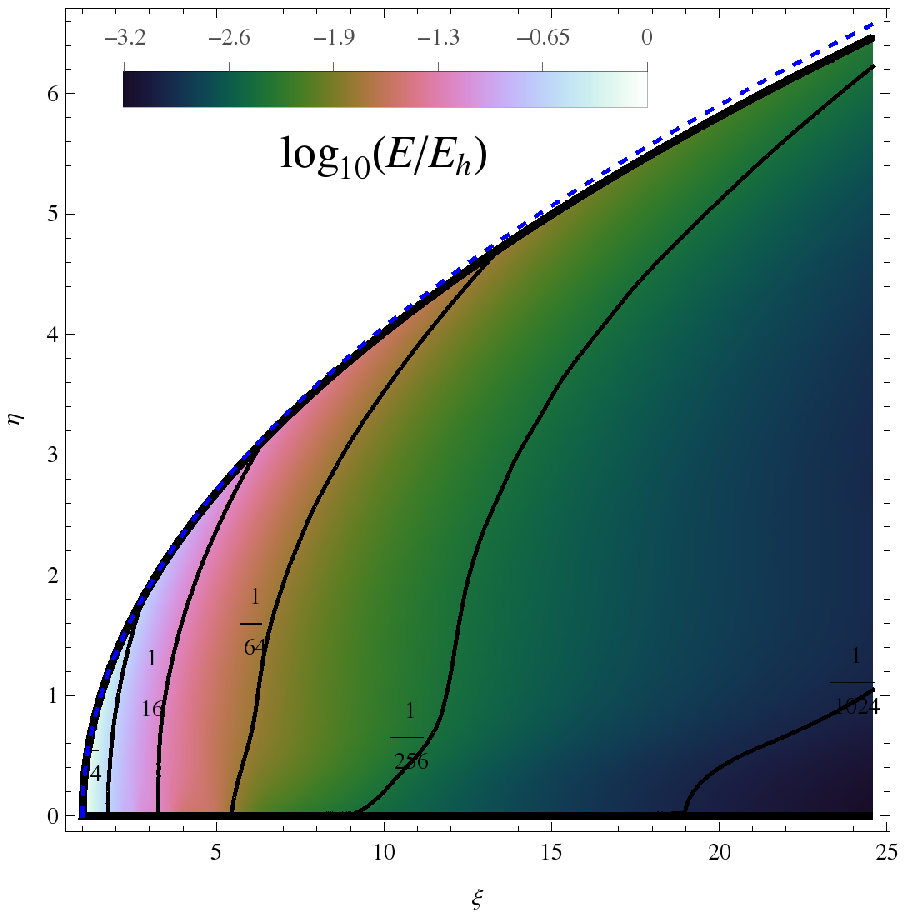}}}
}
\caption{ \label{fig:NumericJet}
Full numerical solution for a high ($n=10$) order expansion of the shock profile.
The difference between the shock front (thick solid curved) and a paraboloid (thick dashed blue) is small, but noticeable.
Far from the head, the minimal $Q$ is found in the envelope at $f\simeq 0.75$ (dotted cyan curve).
Curves with non-monotonic $Q$ first appear at $\xi_Q\simeq 3.8$, roughly located at $f\simeq 0.17$ (dot-dashed red curve); this may be regarded as the boundary between the core and envelope regions.
\BoldA{An inflow,} $U<0$ region appear\BoldA{s} at $\xi_U\simeq 5.5$, confined to $f\lesssim 0.33$ (dot-dashed yellow curve; the narrow $U>0$ stripe along the axis is probably a numerical artifact).
The dot-dashed curves are plotted using Eq.~(\ref{eq:SS_parameterB}), with $\xi_h$ replaced by $\xi_Q$ or $\xi_U$.
\vspace{1cm}
}
% JetSemi/HeadModel36
\end{figure*}

Our convergence tests suggest that the procedure outlined above can in principle be continued indefinitely, to arbitrarily high order, such that a solution in the $n\to\infty$ limit exists.
Namely, given sufficient computational power, the shock profile could be adjusted such that the integration be successfully carried out to arbitrarily large $\xi_{\max}$ and $\rho_{max}$.

The existence of a solution extending infinitely far downstream is further supported by the GSS analysis of \S\ref{sec:GSSequations}--\S\ref{sec:beta202}, in which a solution for the envelope was indeed shown to exist out to $\xi\to\infty$.
However, in the absence of a matched solution for the core, we cannot rigorously prove the existence of the full jet solution much beyond our present, $\xi_{max}\sim 25$ \BoldA{limit}.

Even if we assume that the solution exists, there is no a-priori guarantee that it is physically meaningful, because in realistic scenarios the assumption $q\gg1$ of an ultra-relativistic flow eventually breaks down far downstream. As the downstream flow is subsonic with respect to the shock, the solution could in principle be completely altered once the unavoidable subsonic transition is incorporated.
The physical relevance of the solution would however be assured, if it is a sufficiently strong attractor.

Assuming that the solution exists, it does appear to be unique, because
\myNi
we do not identify any other solution at low order $n\leq 4$, where the phase space can be thoroughly searched for additional solutions, as illustrated in Figure \ref{fig:ParameterScan};
\myNii
such a search is double-checked using the two scanning methods described in \S\ref{sec:NumericalMethod}, found to agree well with each other;
\myNiii
high dimensional scans, although not as complete, show no evidence of other solutions;
and
\myNiv
the semi-analytic constraints of \S\ref{sec:Analytic} limit the phase space of possible solutions.

Numerical simulations (\BoldA{\eg} G07) provide the best evidence that the solution indeed exists, is unique, is physically meaningful, and even behaves as a strong attractor and is likely to be stable, at least against axisymmetric perturbations.
For, in such simulations, roughly tuned for external GRB shock parameters, the self-similar solution is found to emerge from different initial configurations, involving various relativistically moving blobs.

Finally, note that although \BoldA{in \S\ref{sec:equations} and \S\ref{sec:Analytic} we} assumed an infinite jet with $z_s=z_s(t,r)$, the numerical solution \BoldA{here} was derived \BoldA{using} Eq.~(\ref{eq:ShockProfileExpanded}), \BoldA{thus} essentially assuming an $r_s=r_s(t,z)$ shock profile. \BoldA{Hence}, we also search for, and rule out, self-similar jet \BoldA{(\ie relativistic, directed blast wave)} solutions with non-monotonic $\eta_s(\xi)$ profiles, such as pinched jets.

\subsection{Jet structure}
\label{sec:NumericalJetStructure}

The \BoldA{numerical solution to the} self-similar structure of \BoldA{the} jet is shown, for the head region ($\xi<5$) in Figure \ref{fig:Head}, and for the full range available ($1<\xi<\xi_{max}$) in Figure \ref{fig:NumericJet}, using a high ($n=10$) order expansion of the shock profile, optimized in method 1. The corresponding shock profile parameters are listed in Table \ref{tab:shock_profiles}.

As Figure \ref{fig:Head} shows, the numerical results in the head region are fairly well fit by the monotonic head model derived in \S\ref{sec:MonotonicHead} (dashed curves).
It is also in qualitative agreement with the G07 simulations (dot-dashed curves), although the latter correspond to a somewhat narrower shock profile.

In Figure \ref{fig:NumericJet}, the slight deviation of the shock profile from a paraboloid shape (thick dashed curve) becomes apparent far from the head, as $\beta$ approaches its asymptotic value $\beta_\infty\simeq 2.02$.
Indeed, here the envelope of the jet qualitatively resembles the non-monotonic, $\beta=2.02$ GSS profile of figure \ref{fig:Beta202}.

More quantitatively, a radial cross section at large $\xi$ shows the minimum of $Q$ at $f\equiv\eta/\eta_s\simeq 0.75$ (dotted curve), and the sign flip of $U$ at $f\simeq 0.33$ (dot-dashed curve in the $U$ panel).
These values deviate somewhat from those expected for $\beta=2.02$; both would agree with $\beta=2.04$ (see \S\ref{sec:beta202}).
This is not surprising, as our numerical solution only reaches $\eta_{max}\simeq 6.5$, which may not be sufficiently large to show the asymptotic GSS behavior.
Moreover, the $U=0$ contour emerges from the axis only at $\xi=\xi_U\simeq 5.5$; so it may not have converged by $\xi_{max}\simeq 25$.

The breakdown of $Q$ monotonicity is already evident around $\xi=\xi_Q\simeq 4$, and perhaps even closer to the head, but the precise location at which the 'shoulder' in $Q$ (\eg the wiggle in the $Q/Q_h=0.3$ contour shown in Figures \ref{fig:Head} and \ref{fig:NumericJet}) emerges from the axis is not well converged.

At large $\xi$, this feature corresponds to a maximum of $Q$ in a radial cross section, located roughly at $f=f_c\sim 0.17$ (dot-dashed curve in the $Q$ panel of Figure \ref{fig:NumericJet}).
This may be regarded as the boundary between the core and envelope regions.
Note that, although the transition appears to take place at a constant $f$, this did not have to be the case, as the GSS scaling does not apply in the core.

The above results support the partition of the jet into three distinct regions, as anticipated in \S\ref{sec:Analytic}: a head region for $\xi\lesssim \xi_g$, a core region for $\xi_g\lesssim\xi\lesssim\xi_c$, and a GSS envelope for $\xi>\xi_c$.
The boundary of the head region corresponds to the breakdown of monotonicity, $\xi_g\simeq5$, and is related to the onset of non-monotonic $Q$ and negative $U$ behavior, at $\xi_Q$ and $\xi_U$.
The boundary between the core and the envelope can be identified as $f_c\simeq 1/6$, corresponding to $\chi_c\simeq 36$ or equivalently $\xi_c\simeq 20\eta^\beta$.

\section{Summary and discussion}
\label{sec:Discussion}

The equations governing the structure of a self-similar, ultra-relativistic, directed blast wave are derived (Eqs.~\ref{eq:SSt}--\ref{eq:SSBC}), following and correcting G07.
A numerical analysis (\S\ref{sec:NumericalSolution}) suggests a converging (Figure \ref{fig:Convergence1}), unique (Figure \ref{fig:ParameterScan}) jet solution (Figures \ref{fig:Head} and \ref{fig:NumericJet}), which qualitatively agrees with previous simulations (G07) of the head region, and with our semi-analytic study (\S\ref{sec:Analytic}).

The jet can be broadly partitioned into three distinct regimes: a head region ($\xi\lesssim \xi_g\simeq 5$), an axial core ($\xi_g\lesssim\xi\lesssim\xi_c\simeq 20\eta^\beta$), and an envelope ($\xi\gtrsim\xi_c$). The highest Lorentz factors, $\gamma\gtrsim \gamma_h/2$, are found in the head, most of the energy lies in the envelope, and the core contains an axial inflow \BoldA{originating from} the envelope.

In the head region, $Q\gtrsim (Q_h/4)$, $P\gtrsim (P_h/10)$, $U>0$, the shock profile is very close to a paraboloid, and the flow is monotonic, in the sense that $Q$, $P$, and $U$ monotonically decrease away from the shock in both the $\xi$ and the $(-\eta)$ directions.
This region, analyzed in \S\ref{sec:MonotonicHead} and in \S\ref{sec:HeadModel}, is strongly constrained by monotonicity (\eg Figure \ref{fig:HeadMono}); it qualitatively agrees with the head structure reported based on simulations in G07.

The core, analyzed in \S\ref{sec:Axis}, shows a monotonic behavior in $Q$ and $P$.
The radial velocity $U$ is negative here, corresponding to a radial inflow toward the axis; thus $(-U)$, rather than $U$, diminishes toward the axis.
A radial cross section shows $Q$ increasing outward from the axis, until it reaches a maximum at $f\equiv\eta/\eta_s\simeq 1/6$; this marks the transition into the envelope region.

The envelope region follows an additional, geometric self-similarity (GSS), such that the two-dimensional flow in the $\xi$--$\eta$ plane becomes essentially one-dimensional when written in terms of the similarity variable $\chi\sim \xi/\eta^\beta$ (defined more precisely in Eq.~\ref{eq:SS_parameterB}).
Far from the head, \BoldA{the shock power-law index} $\beta = d(\ln\xi_s)/d(\ln\eta)$ asymptotes to $\beta\simeq 2.02$; solving the GSS equations (\ref{eq:GSS_t}--\ref{eq:GSS_BCs}) for this case (Figure \ref{fig:Beta202}) reproduces the envelope structure.
Moreover, the $P$ profile, and the external ($f\gtrsim3/4$; defined by the minimal value of $Q$ in a radial cross section) profiles of $Q$ and $U$, are well approximated by the analytic solution we derive (Eqs.~\ref{eq:beta2approx} and Figure \ref{fig:GSS_beta2}) for the parabolic case $\beta=2$.

The total self-similar energy extrapolated from the numerical solution, $\myenergy_{tot}\simeq 2.1$, fixes the dimensionless constant in Eq.~(\ref{eq:SS_tau}) as $C=\myenergy_{tot}^{-1/3}\simeq 0.8$.
This is about half the value $C=1.5$ estimated in G07 for an initial jet opening angle of $0.1$ radians.
The difference is probably due to the large fraction of the energy deposited far beyond the head \BoldA{region}.
Recovering this energy requires filling the attractor and resolving the far downstream, which is challenging in a numerical simulation.
This incomplete convergence \BoldA{up}on the attractor is likely to be the reason for the differences in detail, seen in Figure \ref{fig:Head}, between \BoldA{our} numerical solution and the G07 simulations.

Our numeric solution appears to be converged, physical, and unique, because
\myNi it \BoldA{qualitatively} agrees with the simulations of G07, as provided for the head region (Figure \ref{fig:Head}), and with the $U$ sign change reported far downstream;
\myNii it agrees qualitatively with the semi-analytic constraints of \S\ref{sec:Analytic};
\myNiii it quantitatively agrees with the GSS envelope solution, which extends to $\xi\to\infty$;
\myNiv convergence tests suggest that it extends to large $\xi$, out to our resolution limit (Table \ref{tab:shock_profiles} and Figure \ref{fig:Convergence1});
and
\myNv no other solution is found in systematic scans of the shock profile at low order (\eg Figure \ref{fig:ParameterScan}), performed in two different methods, nor in high order scans.

Nevertheless, we cannot prove that the solution is an attractor, or that it even survives in physical situations in which the far downstream transitions into the sub-relativistic regime, where our $q\gg1$ assumption fails.
However, the simulations of G07 suggest that a self-similar solution indeed exists, and is a physically relevant, strong attractor.
Under this assumption, our results substantiate the existence of the  solution, uncover its structure, in particular far from the head, and show that it is unique.

Moreover, the G07 simulations then imply that the solution is relevant for the typical parameters of external GRB shocks, and is probably stable, at least under axisymmetric perturbations.
While computing the observational implications is straightforward, and a stability analysis of the self-similar solution appears feasible, we defer th\BoldA{ese} to a forthcoming publication.

\BoldA{
The self-similar regime is strictly valid in the limit of an initially extremely narrow and ultra-relativistic jet, such that causal contact is reached when the shock is still highly relativistic, \ie the opening angle $\theta\lesssim \Gamma^{-1}\ll1$.
For finite initial opening angle $\theta_i$ and Lorentz factor $\Gamma_i$, the attractor is only partly filled.
The simulations of G07 show that explosions with $\theta_i\sim 0.01\till 0.1$ and $\Gamma_i\sim 20$ indeed produce only a partial attractor, gradually filling up throughout the entire quasi-self similar stage.
}

\BoldA{
For GRB afterglow jets, simulations with $\theta_i\gtrsim 0.05$ typically show a slow, non-exponential deceleration and widening \citep{GranotEtAl01, ZhangMacFadyen09, vanEertenEtAl10, MelianiKeppens10, vanEertenMacFadyen11, WygodaEtAl11, DeColleEtAl12}, unlike the self-similar evolution discussed here.
However, a self-similar, exponential expansion is expected for small, $\theta_i\lesssim 0.05$ initial opening angles \citep{WygodaEtAl11, GranotPiran12}, provided that the resolution is sufficiently high \citep{Cannizzo_et_al_04, Granot07, MelianiEtAl07}; such behavior was indeed reported \citep[\eg][and reference therein]{vanEerten13} for jets with $\theta_i\ll0.04$.
}

\acknowledgements

We are most grateful to Ehud Nakar, for extensive discussions and assistance, and to Margaret Pan, for help in the conception of the project.
Ramesh Narayan, Eli Waxman, and Yuri Lyubarsky are acknowledged for useful discussions.
This research has received funding from the European Union Seventh Framework Programme (FP7/2007-2013) under grant agreement n\textordmasculine ~293975 and from an IAEC-UPBC joint research foundation grant, and was supported by the ISF (grant No. 504/14) within the ISF-UGC joint research program framework

%\bibliography{Jet}

\appendix

\section{No real\BoldA{-valued} GSS solution for $\beta<2$}
\label{sec:NoBetaLess2}

Consider the case where $\xi_{s}\propto \eta^{\beta}$ far from the axis, with $\beta<2$.
The GSS scaling is constrained here by the boundary conditions (\ref{eq:SSBC}), giving rise to modified GSS coordinates $\myQb$, $\myPb$, and $\myUb$, defined by
\begin{equation}
Q=A^{-1}\eta^{-\beta}\myQb(\chi) \coma \quad
P=A^{-1}\eta^{-\beta}\myPb(\chi) \coma \quad
U=A\eta^{\beta-1}\myUb(\chi) \coma  \quad
\mbox{and} \quad
\chi\equiv \xi/(A\eta^{\beta}) \coma
\end{equation}
such that Eq.~(\ref{eq:SSBC}) becomes
\begin{equation}
\myQb(1)=\frac{1}{8-4\beta} \coma \quad
\myPb(1)=\frac{1}{6-3\beta} \coma \quad
\mbox{and} \quad
\myUb(1)=\beta \fin
\end{equation}

Far from the axis, the hydrodynamic Eqs.~(\ref{eq:SSt}--\ref{eq:SSr}) now become
\begin{equation}
[1+4(\beta-2)\chi\myQb]\myPb'+4(\beta-2)(2\myQb+\chi \myQb')\myPb=0 \coma
\end{equation}
\begin{equation}
[1+(\beta-2)\chi\myQb]\myQb\myPb'+[(\beta-2)\myQb^{2}-\myQb']\myPb=0 \coma
\end{equation}
and
\begin{equation}
(\myUb-\chi\beta)\myPb'-[\beta-2\myUb'+4(\beta-2)(\myUb-\chi \myUb')\myQb]\myPb=0 \fin
\end{equation}

The first two equations are independent of $\myUb$, and may be solved analytically; $\myQb$ is then given by the transcendental equation
\begin{equation}
[1-(2-\beta)\chi\myQb]^{3}\myQb =\frac{27e^{-1}}{256(2-\beta)}e^{4(2-\beta)\chi\myQb} \fin
\end{equation}
For finite values of $\chi$, this equation has no real-valued solutions for
$\myQb$.

\section{A GSS envelope with $\beta=2$ cannot be matched to the axis}
\label{sec:NoBetaEq2}

For $\beta=2$, the boundary conditions give $U_s\propto\eta$.
In the regime $1/4\leq Q_{a0}<1/2$, this can be matched to an expansion of $U$ along the axis,
\begin{equation} \label{eq:DaniContinuity}
U(\xi,\eta)=c_1\eta+c_3\eta^3\xi^{-1}+c_5\eta^5\xi^{-2}+ \ldots \coma
\end{equation}
where $c_n$ are constants that depend only on $Q_{a0}$.
In this regime, the expansion leads to
\begin{equation}
Q(\xi,\eta)=\xi^{-1}\bar{Q}(\chi)
\quad\mbox{and}\quad
P(\xi,\eta)=\xi^{\frac{-2(2+Q_{a0})}{4+Q_{a0}}}\bar{Q}(\chi) \coma
\end{equation}
where $\bar{Q}$ is an unknown function that satisfies $\bar{Q}(\chi\to\infty)=Q_{a0}$.
As $\chi$ is constant on the shock, $Q$ and $P$ scale differently there, contradicting the boundary conditions (\ref{eq:SSBC}). This rules out a monotonic jet with a $\beta=2$ GSS envelope.

\section{A wide jet leads to a near-head divergence}
\label{sec:WideJets}

The uniqueness of the numerical solution derived in \S\ref{sec:NumericalSolution} suggests that the shock profile is fixed by the regularity of the flow.
To demonstrate how the flow diverges for any deviation from the true shock profile, we consider a wide jet, in which an axial expansion series converges rapidly near the head, in the region where the shock is nearly parabolic, $\xi_s\simeq \xi_h+A\eta^2$.

We expand $Q_a(\xi)$ to orders a high as $n=5$ near the head.
For wide jets with $A\lesssim0.1$, this expansion converges rapidly with $n$ near the shock, and shows that $Q_a$ becomes non-physically zero and subsequently negative near the head.
Moreover, at the point where $Q_a\to0$, $P_a$ and its derivative must also vanish, and $U_1$ diverges.
This vanishing of $Q$ near the head of a wide jet is verified numerically, and traced even for $A\lesssim 0.3$.

\section{Monotonic head model}
\label{sec:HeadModel}

An approximate, monotonic head description may be found using the axial expansion (\ref{eq:Qexpansion}--\ref{eq:Uexpansion}), truncated beyond order $\eta^5$.
A monotonic profile far from the head requires $Q_a(\xi\gg1)\propto \xi^{-1}$ (see \S\ref{sec:Axis}).
For simplicity, we assume that $Q_a\propto \xi^{-1}$ even near the head, so the boundary conditions fix $Q_a=1/(8\xi)$.
The axial analysis then implies that $P_a=(1/6)\xi^{-5/3}$.
The axial analysis also implies that $U_1(\xi)=1$, but it is more accurate to determine \BoldA{$U_1$} directly from the shock boundary conditions, as shown below.

Combining Eqs.~(\ref{eq:SSt}--\ref{eq:SSBC}) and (\ref{eq:ShockDerivative}), the expansion coefficients of $P$ and $Q$ are fixed up to order $\eta^4$, and the coefficients of $U$ are fixed up to order $\eta^5$, by the shock boundary conditions on $Q$, $P$ and $U$, their first derivatives, and the second derivative of $U$.

This can be done for any shock profile, but the resulting coefficient expressions are in general lengthy.
For brevity, we provide the expansion coefficients for the parabolic profile $\xi_{s}=\xi_{h}+A\eta^{2}$, accurate close to the nose of the jet.
Here, the boundary conditions are explicitly given by
\begin{equation}
Q_{s}=\frac{1}{8A^{2}\eta^{2}+8\left(A\eta^{2}+\xi_{h}\right)-8A\eta^{2}}
\quad\mbox{and}\quad
Q'_{s} = \frac{A\eta\left(A^{3}\eta^{2}-3A\xi_{h}+2\xi_{h}\right)}{4\left(A^{2}\eta^{2}+\xi_{h}\right)^{3}} \coma
\end{equation}
\begin{equation}
P_{s}=\frac{4}{3}Q_{s}\quad\mbox{and}\quad
P'_{s}  =  \frac{7A\eta\left(A^{3}\eta^{2}-A\xi_{h}+\xi_{h}\right)}{9\left(A^{2}\eta^{2}+\xi_{h}\right)^{3}}
 \coma
\end{equation}
and
\begin{equation}
U_{s}=2A\eta \coma\quad
U'_{s}=2A \coma\quad\mbox{and}\quad
U''_{s}=-\frac{32A^{4}\eta^{3}\left(A^{3}\eta^{2}-3A\xi_{h}+2\xi_{h}\right)}{3\left(A^{2}\eta^{2}+\xi_{h}\right)^{3}} \fin
\end{equation}
The expansion coefficient solutions are then
\begin{flushleft}
\begin{flalign}
Q_{2}(\xi) = & \frac{(1-2A)A^{3}\xi^{2}+A(A(A(4A-7)+7)-2)\xi\xi_{h}-2A(A-1)^{3}\xi_{h}^{2}}{8\xi(A(\xi-\xi_{h})+\xi_{h})^{3}} \coma \nonumber
\\
Q_{4}(\xi)= &\frac{A^{2}\left(A^{3}\xi^{2}+(A((3-2A)A-5)+2)\xi\xi_{h}+(A-1)^{3}\xi_{h}^{2}\right)}{8\xi(\xi-\xi_{h})(A(\xi-\xi_{h})+\xi_{h})^{3}}
 \coma \nonumber \\
P_{2}(\xi) = & \frac{A\left(-6A^{3}(\xi-\xi_{h})^{3}-A^{2}\left(\xi^{5/3}+18\xi_{h}\right)(\xi-\xi_{h})^{2}+A\xi_{h}\left(19\xi^{5/3}-18\xi_{h}\right)(\xi-\xi_{h})\right)}{18\xi^{5/3}(\xi-\xi_{h})(A(\xi-\xi_{h})+\xi_{h})^{3}} \nonumber \\
 & + \frac{\text{A\ensuremath{\xi}h}\left(13\xi^{5/3}\xi_{h}-7\xi^{8/3}-6\xi_{h}^{2}\right)}{18\xi^{5/3}(\xi-\xi_{h})(A(\xi-\xi_{h})+\xi_{h})^{3}} \coma \nonumber \\
P_{4}(\xi) = & \frac{A^{2}\left(3A^{3}(\xi-\xi_{h})^{3}+A^{2}\left(4\xi^{5/3}+9\xi_{h}\right)(\xi-\xi_{h})^{2}-A\xi_{h}\left(13\xi^{5/3}-9\xi_{h}\right)(\xi-\xi_{h})\right)}{18\xi^{5/3}(\xi-\xi_{h})^{2}(A(\xi-\xi_{h})+\xi_{h})^{3}} \nonumber \\
 & + \frac{A^{2}\xi_{h}\left(-10\xi^{5/3}\xi_{h}+7\xi^{8/3}+3\xi_{h}^{2}\right)}{18\xi^{5/3}(\xi-\xi_{h})^{2}(A(\xi-\xi_{h})+\xi_{h})^{3}} \coma \nonumber \\
U_{1}(\xi)= & \frac{2A\left(A^{3}(\xi-\xi_{h})^{3}+15A^{2}\xi_{h}(\xi-\xi_{h})^{2}-A\xi_{h}(4\xi-13\xi_{h})(\xi-\xi_{h})+3\xi_{h}^{3}\right)}{3(A(\xi-\xi_{h})+\xi_{h})^{3}} \coma \nonumber \\
U_{3}(\xi)= & \frac{8A^{3}(\xi-\xi_{h})(A(A\xi-(A+3)\xi_{h})+2\xi_{h})}{3(A(\xi-\xi_{h})+\xi_{h})^{3}} \coma \nonumber \\
\mbox{and} \quad \quad\quad
U_{5}(\xi)= &\frac{4A^{4}(A(A+3)-2)\xi_{h}-4A^{6}\xi}{3(A(\xi-\xi_{h})+\xi_{h})^{3}} \fin &&
\end{flalign}
\end{flushleft}

\end{document}